\documentclass[aps,prx,amsmath,amssymb,longbibliography,lengthcheck,showpacs,superscriptaddress,floatfix]{revtex4-1}
\usepackage{graphicx}

\begin{document}

\title{Phonon transport and vibrational excitations in amorphous solids}

\author{Hideyuki Mizuno}
\email{hideyuki.mizuno@phys.c.u-tokyo.ac.jp} 
\affiliation{Graduate School of Arts and Sciences, The University of Tokyo, Tokyo 153-8902, Japan}

\author{Atsushi Ikeda}
\affiliation{Graduate School of Arts and Sciences, The University of Tokyo, Tokyo 153-8902, Japan}

\date{\today}

\begin{abstract}
One of the long-standing issues concerning the thermal properties of amorphous solids is the complex pattern of phonon transport.
Recent advances in experiments and computer simulations have indicated a crossover from Rayleigh scattering to $\Omega^2$ law (where $\Omega$ is the propagation frequency).
A number of theories have been proposed, yet critical tests are missing and the validity of these theories is unclear.
In particular, the precise location of the crossover frequency remains controversial, and more critically, even the validity of the Rayleigh scattering itself has been seriously questioned.
To settle these issues, we focus on a model amorphous solid, whose vibrational eigenmodes were recently clarified over a wide frequency regime: a mixture of phonon modes and soft localized modes in the continuum limit and disordered and extended modes in the boson peak regime.
The present work demonstrates that Rayleigh scattering occurs in the continuum limit and $\Omega^2$ damping occurs in the boson peak regime, and these behaviors are therefore linked to the underlying eigenmodes in the corresponding frequency regimes.
Our results unambiguously determine the crossover frequency.
Furthermore, we establish characteristic scaling laws of phonon transport near the jamming transition, which are consistent with the prediction of the mean-field theory at higher frequencies but inconsistent in the low-frequency, Rayleigh scattering regime.
Our results therefore reveal crucial issues to be solved with regard to the current theory.
\end{abstract}

\maketitle

\section{Introduction}
An understanding of the low-temperature thermal properties of crystalline solids is well established~\cite{Ashcroft,Leibfried}.
The heat capacity $C$ and the thermal conductivity $\kappa$ follow the universal laws of $C \propto T^3$ and $\kappa \propto T^3$~(where $T$ is the temperature), which are explained by the Debye theory and the phonon gas theory, respectively.
In contrast, amorphous solids show thermal properties that are markedly different from those of crystalline solids~\cite{lowtem,Zeller_1971,Graebner_1986}.
The heat capacity increases linearly with temperature, $C \propto T$, at $T \lesssim 1$~[K].
As $T$ increases, $C$ approaches the $T^3$ dependence of the Debye law, but the reduced heat capacity $C/T^3$ exhibits a peak at $T \sim 10$~[K], which is called the boson peak~(BP).
In addition, the thermal conductivity increases as $\kappa \propto T^2$ at $T \lesssim 1$~[K] and exhibits a plateau at $T \sim 10$~[K].
These anomalous temperature dependences are surprisingly universal; they are shared by many different amorphous solids, regardless of their constituents~\cite{lowtem,Zeller_1971,Graebner_1986}.

The anomalies in the thermal properties of amorphous solids originate from anomalies in their vibrational properties.
The BP in the heat capacity indicates that the vibrational density of states (vDOS) $g(\omega)$~(where $\omega$ is the frequency) does not follow the Debye law, $g(\omega) \propto \omega^2$.
In fact, many amorphous solids universally exhibit a peak at $\omega = \omega_\text{BP} \sim 1$~[THz] in the reduced vDOS $g(\omega)/\omega^2$~\cite{buchenau_1984,Yamamuro_1996,Kabeya_2016}; this peak is also called the BP.
The complex temperature dependence of the thermal conductivity indicates that the phonon transport in these materials is very different from that in crystalline solids.
Recent advanced experiments~\cite{ruffle_2006,Masciovecchio_2006,Monaco_2009,Baldi_2010,Baldi_2011,Baldi_2016} have observed that the sound speed exhibits a drop near the BP frequency, at $\Omega \sim \omega_\text{BP}$ (where $\Omega$ is the propagation frequency), which is referred to as sound softening.
They have also reported that the sound attenuation rate $\Gamma$ depends on $\Omega$ in a complex manner.
When $\Gamma$ is fitted to a power law, $\Gamma \propto \Omega^n$, the exponent $n$ depends on the $\Omega$ regime: $n \approx 2$ (anharmonic damping) at low $\Omega \ll \omega_\text{BP}$, $n \approx 4$ (Rayleigh scattering) at intermediate $\Omega \lesssim \omega_\text{BP}$, and $n \approx 2$ ($\Omega^2$ law) at high $\Omega \gtrsim \omega_\text{BP}$.
Understanding these vibrational properties is of key importance because they explain the thermal properties of amorphous solids.

So far, a number of theoretical explanations have been proposed, which are qualitatively different from each other.
One approach is based on the proposed concept of elastic heterogeneities~\cite{schirmacher_2006,schirmacher_2007,Marruzzo_2013,Schirmacher_2015,Monaco2_2009,Mizuno2_2013,Mizuno_2014,Gelin_2016,Zhang_2017}.
In this approach, spatially fluctuating local elastic moduli~\cite{yoshimoto_2004,Wagner_2011,Mizuno_2013} give rise to the vibrational properties of amorphous solids and, hence, their thermal properties.
A second approach is based on a two-level system and a soft potential model~\cite{Anderson_1972,Karpov_1983,Buchenau_1991,Buchenau_1992,Gurevich_2003,Gurevich_2005,Gurevich_2007}.
In this approach, two-level systems~\cite{Heuer_1993,Heuer_1996,Perez-Castaneda_2014,Perez-Castaneda2_2014,Damart_2018} and soft localized vibrations~\cite{mazzacurati_1996,Taraskin_1999,Xu_2010,Baity-Jesi2015,Beltukov_2016,Lerner_2016} are essential ingredients in producing the vibrational and thermal anomalies.
A third approach is based on the mean-field theory of jamming~\cite{Wyart_2005,Wyart_2006,Silbert_2005,Silbert_2009,Xu_2009,Vitelli_2010,Vitelli2_2010,Parisi2010,Wyart_2010,Berthier2011a,DeGiuli_2014,Charbonneau2014b,Franz_2015,Milkus_2016,Charbonneau_2016,Biroli_2016}, in which isostaticity and marginal stability are crucial.
It is not easy to judge the validity of these theories because critical tests are difficult to conduct due to the presence of phenomenological fitting parameters.
In this regard, an exception might be the mean-field theory of jamming, as its predictions regarding scaling laws can be precisely tested, as we will discuss in detail. 

Numerical simulations would be an ideal tool for studying vibrational properties because they enable us to directly investigate the vibrations of particles.
However, this is not as easy as it sounds because one needs to analyze large systems in order to access a wide range of frequencies, particularly the low-frequency regime.
Only recently have large-scale simulations of amorphous solids been performed on, which have enabled progress in the understanding of their vibrational properties.
As we will describe in the Preliminaries section, the vDOS and the vibrational eigenmodes have been elucidated over a vast frequency range.
In particular, our previous works~\cite{Mizuno_2017,Shimada_2017} observed that phonon modes coexist with soft localized modes in the low-frequency regime.
The nature of these soft localized modes was studied in detail in Refs.~\cite{Lerner_2016,Gartner_2016,Lerner_2017,Lerner2_2017}.
In addition, a recent work~\cite{Bouchbinder_2018} reported that the soft localized modes can hybridize with the phonon modes as the system size increases.
(We will discuss effects of the hybridization in the Conclusion section.)

On the other hand, phonon transport has been less extensively studied and remains controversial.
It has been reported that the sound speed exhibits a drop at $\Omega \sim \omega_\text{BP}$ and that the attenuation rate shows a crossover at $\Omega \sim \omega_\text{BP}$ from Rayleigh scattering, $\Gamma \propto \Omega^4$, to $\Omega^2$ law, $\Gamma \propto \Omega^2$~\cite{Monaco2_2009,Marruzzo_2013,Mizuno_2014}.
Although these results are consistent with experimental observations~\cite{ruffle_2006,Masciovecchio_2006,Monaco_2009,Baldi_2010,Baldi_2011,Baldi_2016}, the precise location of the crossover frequency has not yet been settled.
Moreover, the length scale of the structural disorder responsible for the Rayleigh scattering remains unclear~\cite{Vitelli2_2010}.
Furthermore, and more critically, even the validity of the Rayleigh scattering law has been seriously questioned in recent work~\cite{Gelin_2016}.
The authors of Ref.~\cite{Gelin_2016} claimed that both the numerical results and the experimental datasets can be fitted to $\Gamma \propto -q^{d+1} \log q$ (where $q$ is the wavenumber and $d$ is the spatial dimension).
They also argued that the logarithmic correction to the Rayleigh law might originate from the long-range nature of elastic disorder~\cite{DeGiuli_2018}.

The present work aims to settle all of these issues and establish a comprehensive understanding of the phonon transport in amorphous solids.
To this end, we focus on randomly packed soft particles near the jamming transition as a model of an amorphous solid~\cite{OHern_2003} and perform large-scale numerical simulations. 
We accomplish the following two goals. 
(1) We capture the complex phonon transport properties of amorphous solids in simulations. 
We then explain these phonon transport properties in terms of the underlying vibrational eigenmodes. 
This is possible because we have already revealed the nature of the eigenmodes of the present model~\cite{Mizuno_2017}. 
(2) We establish the scaling laws of the phonon transport properties near the jamming transition. 
This enables us to directly test (without any phenomenological fitting parameters) the predictions of the mean-field theory of jamming~\cite{Wyart_2010,DeGiuli_2014}. 
We show that the scaling law of the mean-field theory works well at higher frequencies but breaks down in the low-frequency, Rayleigh scattering regime.

\section{Preliminaries}
We first review some basic features of the vibrational eigenmodes in the present model of an amorphous solid.
Our model is composed of randomly jammed particles interacting via the following finite-range, purely repulsive potential~\cite{OHern_2003}:
\begin{equation}~\label{pot-simple}
\phi(r) = \frac{\epsilon}{2} \left( 1 - \frac{r}{\sigma} \right)^2 H(\sigma -r),
\end{equation}
where $\sigma$ is the particle diameter and $H(r)$ is the Heaviside step function (see also Section~\ref{system}).
This model exhibits a jamming transition at the packing fraction $\varphi_J$, at which the particles lose their contacts and rigidity.
We note that $\varphi_J \approx 0.64$ and~$\approx 0.84$ in three-dimensional (3D) and two-dimensional (2D) spaces, respectively.
Near the jamming transition, the packing pressure scales as $p \propto (\varphi - \varphi_J)$.
In this work, we employ the pressure $p$ as the control parameter.

Previous numerical studies~\cite{Silbert_2005,Silbert_2009,Charbonneau_2016,Mizuno_2017} have established that the vDOS of this model obeys the following scaling laws in a 3D space:
\begin{equation} \label{vdos}
g(\omega) = 
\left\{ 
\begin{aligned}
& \alpha_\ast & (\omega \gtrsim \omega_\ast ), \\
& \alpha_\text{BP} \left( \frac{\omega}{\omega_\ast} \right)^2 & (\omega \sim \omega_\text{BP}), \\
& A_0 \omega^2 + \alpha_\text{loc} \left( \frac{\omega}{\omega_\ast} \right)^4 & (\omega \lesssim \omega_\text{ex0}),
\end{aligned} 
\right.
\end{equation}
where $\alpha_\ast$, $\alpha_\text{BP}$, and $\alpha_\text{loc}$ are constants and $A_0$ is the Debye level.
Note that $A_0$ scales with $p$ as $A_0 \propto p^{-3/4}$ in the 3D case.
The three characteristic frequencies, $\omega_\text{ex0},$ $\omega_\text{BP}$, $\omega_\ast$ ($\omega_\text{ex0} < \omega_\text{BP} < \omega_\ast$), follow the scaling law $\omega_\text{ex0} \propto \omega_\text{BP} \propto \omega_\ast \propto p^{1/2}$. 
(We explicitly plot these characteristic frequencies as functions of $p$ in Fig.~\ref{figs6a}.)
The nature of the vibrational eigenmodes varies as $\omega$ changes~\cite{Mizuno_2017}: 
(i) At high $\omega > \omega_\ast$, the vibrational modes share similarities with floppy modes and are characterized by disordered and extended vibrations. 
(ii) As $\omega$ decreases, the reduced vDOS $g(\omega)/\omega^2$ exhibits a BP at $\omega \sim \omega_\text{BP}$.
The vibrational modes are similar to those at $\omega > \omega_\ast$ but show more phonon-like characteristics.
(iii) At low $\omega \lesssim \omega_\text{ex0}$, the vibrational modes split into two types: phonon modes and soft localized modes
\footnote{
The phonon modes are not exactly phonons but rather are weakly perturbed.
Exact phonons may be realized only in the zero-frequency limit, $\omega \to 0$.
}.
The phonon modes follow the Debye law, $g(\omega) = A_0 \omega^2$, whereas the soft localized modes follow a non-Debye law, $g(\omega) \propto \omega^4$.
We refer to this frequency regime ($\omega \lesssim \omega_\text{ex0}$) as the continuum limit
\footnote{
Recently it was argued that the soft localized modes would hybridize with the phonon modes when the system is large enough~\cite{Bouchbinder_2018}.
If this is the case, the definition of $\omega_\text{ex0}$ would need a caution because observation of the split of modes is not straightforward.
However, we would consider that $\omega_\text{ex0}$ can be well defined even in such the situations, and at $\omega_\text{ex0}$ the nature of vibrational eigenmodes qualitatively changes.
This point is discussed in the Conclusion section.
}.

\section{Numerical methods}
\subsection{System description}~\label{system}
The present systems are the same as those studied in our previous work~\cite{Mizuno_2017}: a three-dimensional (3D, $d=3$) model and a two-dimensional (2D, $d=2$) model of amorphous solids composed of randomly jammed soft particles.
Particles $i$ and $j$ interact via a finite-range, purely repulsive, harmonic potential, $\phi(r)$, as given in Eq.~(\ref{pot-simple}).
The 3D system consists of monodisperse particles with a diameter of $\sigma$, whereas the 2D system is a $50\%$-$50\%$ binary mixture with a size ratio of $1.4$ (where the diameter of the smaller species is denoted by $\sigma$).
The particle mass is $m$.
Length, mass, and time are measured in units of $\sigma$, $m$, and $\tau=(m \sigma^2/\epsilon)^{1/2}$, respectively.

The control parameter of this system is the packing pressure $p$.
We note that the temperature $T$ is held at zero, $T \equiv 0$.
This amorphous system exhibits a(n) (un)jamming transition~\cite{OHern_2003}: as $p$ decreases, the particles lose their contacts at $p=0$.
In the vicinity of the jamming transition, the system is governed by various power-law scalings with $p$ for quantities such as the elastic moduli $K$ and $G$ and the frequency $\omega_\ast$~\cite{OHern_2003,Silbert_2005}.
We prepared the system with a wide range of $p = 5\times 10^{-2}$ to $1 \times 10^{-4}$, as done in our previous work~\cite{Mizuno_2017}.

To access the low-frequency regime, including the continuum limit $\omega < \omega_\text{ex0}$ ($\omega < \omega_0$), we employed large system sizes: $N=1024000$, $2048000$, and $4096000$ for the 3D case and $N=1024000$ for the 2D case.
(We denote the frequency of continuum limit as $\omega_0$ in the 2D case~\cite{Mizuno_2017}.)
Here, we always removed rattler particles, namely, those in contact with fewer than $d$ other particles.
Note that the present work studies harmonic vibrational properties, where the rattler particles play no roles.
However, we expect that the rattlers make some impacts on anharmonic vibrational properties.

In the present system, the interparticle forces are always positive, $-\phi'(r) > 0$.
For this reason, we refer to this original system as the stressed system.
In addition to the stressed system, we also study the unstressed system.
In the unstressed system, we retain the stiffness $\phi''(r)$ but drop the forces, setting $-\phi'(r) \equiv 0$ in the analysis.
Since the positive forces make the system mechanically unstable, dropping the forces causes the originally stressed system to become more stable~\cite{Wyart_2006,DeGiuli_2014,Lerner_2014}.
We note that in the unstressed system, $\omega_\ast$ coincides with the BP frequency, $\omega_\ast = \omega_\text{BP}$.
In addition, the soft localized modes are significantly suppressed in the unstressed system~\cite{Mizuno_2017,Lerner2_2017}.

\begin{figure}[t]
\centering
\includegraphics[width=0.475\textwidth]{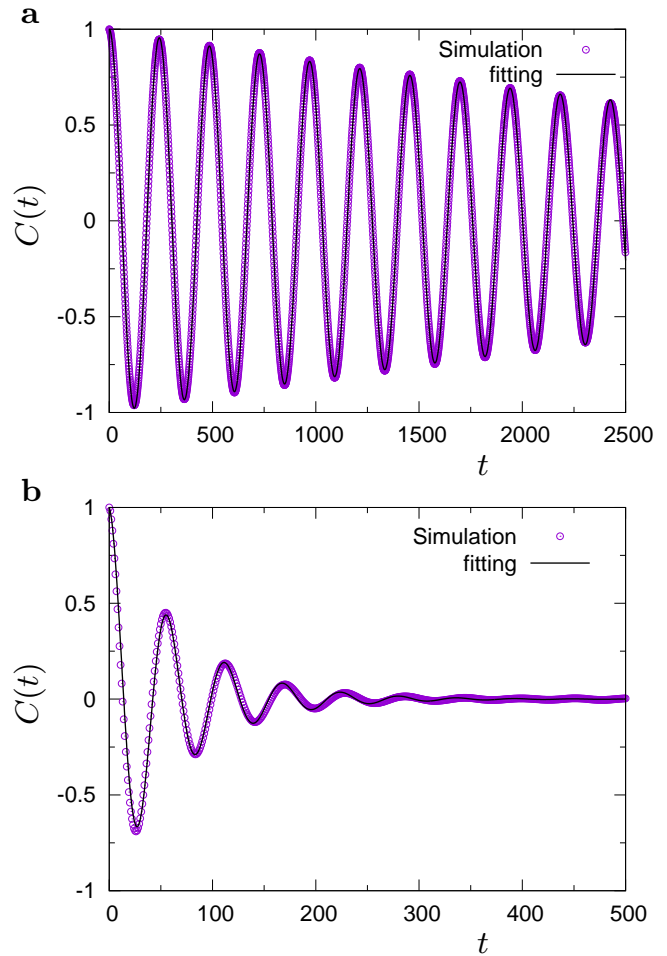}
\vspace*{0mm}
\caption{\label{figs8}
{Time evolution of the velocity-velocity correlation function.}
Plots of the correlation function $C(t)$ as a function of $t$, for transverse waves ($\alpha = T$).
The system is the 3D model system.
{\bf a}, $q= 0.088$, $\Omega = 0.026$ ($< \omega_\text{ex0}$), and $\Gamma_T = 3.8 \times 10^{-4}$.
{\bf b}, $q= 0.39$, $\Omega = 0.11$ ($\approx \omega_\text{BP}$), and $\Gamma_T = 3.0 \times 10^{-2}$.
The symbols represent simulation data.
To quantify the frequency $\Omega$ and the attenuation rate $\Gamma_T$, the simulation data were fitted with the functional form $C(t) \equiv \cos(\Omega t) e^{-\Gamma_\alpha t/2}$; the fitting results are shown as solid lines.
}
\end{figure}

\subsection{Phonon transport analysis}~\label{methodphonon}
To measure the phonon transport, we employed the numerical simulation method used in Ref.~\cite{Gelin_2016}, in which the vibrational dynamics around the equilibrium configuration (inherent structure) are analyzed within the harmonic approximation.
Below, we denote the position of particle $i$ ($i=1,2,3,...,N$) in the inherent structure by $\mathbf{r}_i$ and the displacement from $\mathbf{r}_i$ by $\mathbf{u}_i$.

We first excite a phonon at the initial time $t=0$ by perturbing the velocity $\dot{\mathbf{u}}_i$ of particle $i$:
\begin{equation}
\dot{\mathbf{u}}_i(t=0) = \dot{\mathbf{u}}^0_i \equiv {\mathbf{a}}_\alpha \sin \left( \mathbf{q} \cdot \mathbf{r}_i + \psi \right),
\end{equation}
where $\mathbf{q}$ is the wave vector; $q \equiv \left| \mathbf{q} \right|$ is the wavenumber; $\alpha$ denotes the polarization, with $\alpha = L$ indicating longitudinal waves and $\alpha = T$ indicating transverse waves; and $\psi$ is set to $0$ or $\pi/2$.
The polarization vector (unit vector) ${\mathbf{a}}_\alpha$ is determined as follows: $\mathbf{a}_L = {\mathbf{q}}/q$ for the longitudinal case and $\mathbf{a}_T \cdot {\mathbf{q}} = 0$ for the transverse case.

We next solve the linearized equation of motion:
\begin{equation}
\ddot{\mathbf{u}}_i = \sum_{j=1}^{N} \mathbf{D}_{ij} \cdot \mathbf{u}_j + \dot{\mathbf{u}}^0_i \delta (t),
\end{equation}
where $\mathbf{D}_{ij}$ is the dynamical matrix~\cite{Ashcroft,Leibfried}, and $\delta(t)$ is the Dirac delta function.
From the time history $\mathbf{u}_i(t)$, we calculate the normalized velocity-velocity correlation function:
\begin{equation}
C(t) \equiv \frac{ {\left( \sum_{i=1}^{N} \dot{\mathbf{u}}_i(t) \cdot \dot{\mathbf{u}}_i^0 \right)} }{ {\left( \sum_{i=1}^{N} \dot{\mathbf{u}}^0_i \cdot \dot{\mathbf{u}}^0_i \right)} }.
\end{equation}
The function $C(t)$ represents the propagation and attenuation behaviors of the initially excited phonon $\dot{\mathbf{u}}^0_i$.

We performed repeated simulations of a phonon $\dot{\mathbf{u}}^0_i$ with a wavenumber $q$ and a polarization $\alpha$ with the following settings: $\mathbf{q}=(q,0,0)$, $(0,q,0)$, and $(0,0,q)$ in the 3D case or $\mathbf{q}=(q,0)$ and $(0,q)$ in the 2D case and $\psi=0$ and $\pi/2$.
In addition, for the 3D case, there are two transverse waves with different polarization vectors, $\mathbf{a}_{T_1}$ and $\mathbf{a}_{T_2}$, which were also simulated independently.
The final value of $C(t)$ for a given $q$ and $\alpha$ was obtained by averaging over these cases.
Note that since we implemented periodic boundary conditions in all directions, $q$ takes discrete values of $q=(2\pi/L)n$, where $L$ is the system size and $n=1,2,3,...$ is an integer.

We finally fit the functional form of 
\begin{equation}~\label{ctfunction}
C(t) \equiv \cos(\Omega t) e^{-\Gamma_\alpha t/2},
\end{equation}
to the simulation data and quantified the propagation frequency $\Omega$, the sound speed $c_\alpha \equiv \Omega/q$, and the attenuation rate $\Gamma_\alpha$ (see Fig.~\ref{figs8}).
The scattering length was obtained as $\ell_{\alpha} = 2c_\alpha / \Gamma_\alpha$.
We note that the values of $\Omega$, $c_\alpha$, and $\Gamma_\alpha$ are functions of the wavenumber $q$; alternatively, we can treat $c_\alpha$ and $\Gamma_\alpha$ as functions of $\Omega$ by transforming $q$ into $\Omega$ via the relation $\Omega = \Omega(q)$.

\subsection{Generalized Debye model}~\label{sectionGDM}
In this section, we formulate the vDOS in the framework of the generalized Debye model~\cite{schirmacher_2006,schirmacher_2007,Marruzzo_2013,Schirmacher_2015}.
In this framework, we assume the phonon approximation, $\Omega \gg \Gamma_\alpha$, which is shown to be valid in the low-frequency regime $\Omega \lesssim \omega_\text{BP}$ below the BP (see Appendix~\ref{IRsection}).

The Green function is defined as~\cite{Leibfried}
\begin{equation}
G(t) \equiv \frac{ \left( \sum_{i=1}^{N} {\mathbf{u}}_i(t) \cdot \dot{\mathbf{u}}_i^0 \right) }{ \left( \sum_{i=1}^{N} \dot{\mathbf{u}}^0_i \cdot \dot{\mathbf{u}}^0_i \right) } = \int C(t) dt.
\end{equation}
Using the functional form of $C(t)$ in Eq.~(\ref{ctfunction}), we obtain
\begin{equation}
G(t) \approx  \left\{ \frac{\sin(\Omega t)}{\Omega} \right\} e^{-\Gamma_\alpha t/2} H(t),
\end{equation}
where $H(t)$ is the Heaviside step function.
The Fourier transform of $G(t)$ is formulated as
\begin{equation} \label{greenf2}
\tilde{G}_\alpha(q,\omega) = \int_{-\infty}^{+\infty} G(t) e^{i\omega t} dt \approx \frac{1}{-\omega^2 + q^2 \hat{c}_\alpha(q,\omega)^2},
\end{equation}
where $\hat{c}_\alpha(q,\omega)$ is the complex sound speed:
\begin{equation}
\hat{c}_\alpha(q,\omega) = c_\alpha(q) \left\{ 1 - i \frac{\omega \Gamma_\alpha(q)}{\Omega(q)^2} \right\}^{1/2}.
\end{equation}
The $\omega$-functional form of $\tilde{G}_\alpha(q,\omega)$ in Eq.~(\ref{greenf2}) is the damped harmonic oscillator model, which has been employed in many previous works~\cite{shintani_2008,Monaco2_2009,Baldi_2011,Mizuno_2014,Baldi_2016}.
We also note that the present simulation method~\cite{Gelin_2016} analyzes the Green function that is considered in the relevant theories~\cite{schirmacher_2006,schirmacher_2007,Marruzzo_2013,Schirmacher_2015,Wyart_2010,DeGiuli_2014}, \textit{not} the dynamical structure factor $S(q,\omega)$ that is accessible through scattering experiments~\cite{ruffle_2006,Masciovecchio_2006,Monaco_2009,Baldi_2010,Baldi_2011,Baldi_2016}.
From $S(q,\omega)$, we measure the resonant sound speed $c_{\alpha \text{res}}$ and the attenuation rate $\Gamma_{\alpha \text{res}}$.
However, as long as we focus on the low-$\Omega$ excitations and the phonon approximation is valid, we can expect $c_\alpha \simeq c_{\alpha \text{res}}$ and $\Gamma_\alpha \simeq \Gamma_{\alpha \text{res}}$.
This point was discussed in detail in Ref.~\cite{DeGiuli_2014}.

Following Ref.~\cite{Marruzzo_2013}, we approximate $\hat{c}_\alpha(q,\omega)$ by dropping the dependence on the wavenumber $q$, as follows:
\begin{equation}
\begin{aligned}
\hat{c}_\alpha(q,\omega) &\approx \hat{c}_\alpha \left(q=\Omega^{-1}(\omega),\omega \right), \\
 &= c_\alpha(\omega) \left\{ 1 - i \frac{\Gamma_\alpha(\omega)}{\omega} \right\}^{1/2},
\end{aligned}
\end{equation}
where $\Omega^{-1}$ denotes the inverse function, $c_\alpha(\omega) \equiv c_\alpha(q=\Omega^{-1}(\omega))$, and $\Gamma_\alpha(\omega) \equiv \Gamma_\alpha(q=\Omega^{-1}(\omega))$.
This approximation is valid in the phonon approximation, $\Omega(q) \gg \Gamma_\alpha (q)$.
Note that a recent experiment~\cite{Baldi_2016} tested this approximation and confirmed its validity below the BP frequency.
We then formulate $\tilde{G}_\alpha(q,\omega)$ as
\begin{equation} \label{greenf3}
\tilde{G}_\alpha(q,\omega) \approx \frac{1}{ \left\{ 1 - i \frac{\Gamma_\alpha(\omega)}{\omega} \right\} \left\{ -\omega^2 + q^2 c_\alpha(\omega )^2 \right\} }.
\end{equation}

By using the Green function $\tilde{G}_\alpha(q,\omega)$, the vDOS can be formulated as
\begin{equation}
\begin{aligned}
g(\omega) &= \left({2\omega}/{\pi q_D^d} \right) \\
& \ \times \int_{0}^{q_D} dq q^{d-1} \text{Im} \left\{ (d-1) \tilde{G}_T({q},\omega) + \tilde{G}_L({q},\omega) \right\},
\end{aligned}
\end{equation}
where $q_D = \sqrt[d]{2d\pi^{d-1} \hat{\rho}}$ ($\hat{\rho}=N/L^d$ is the number density) is the Debye wavenumber.
Using $\tilde{G}_\alpha(q,\omega)$ in Eq.~(\ref{greenf3}), we derive the following for the 3D case ($d=3$):
\begin{equation}
\begin{aligned} \label{3dpvdos}
& g(\omega) = \left( \frac{2}{c_T^3 q_D^3} + \frac{1}{c_L^3 q_D^3} \right) \omega^2 \\
& + \frac{2}{\pi} \Bigg[
\frac{2\Gamma_T}{c_T^2 q_D^2} \left\{ 1 + \left( \frac{\omega}{2c_T q_D}\right) \log\left( \frac{ c_T q_D-\omega}{c_T q_D+\omega} \right) \right\} \\
& +
\frac{\Gamma_L}{c_L^2 q_D^2} \left\{ 1 + \left( \frac{\omega}{2c_L q_D}\right) \log\left( \frac{ c_L q_D-\omega}{c_L q_D+\omega} \right) \right\}
\Bigg].
\end{aligned}
\end{equation}
Similarly, we derive the following for the 2D case ($d=2$):
\begin{equation}
\begin{aligned} \label{2dpvdos}
& g(\omega) = \left( \frac{1}{c_T^2 q_D^2} + \frac{1}{c_L^2 q_D^2} \right) \omega \\
& + \frac{1}{\pi} \Bigg\{
\frac{\Gamma_T}{c_T^2 q_D^2} \log\left( \frac{c_T^2 q_D^2}{\omega^2} - 1 \right) + \frac{\Gamma_L}{c_L^2 q_D^2} \log\left( \frac{c_L^2 q_D^2}{\omega^2} - 1 \right)
\Bigg\}.
\end{aligned}
\end{equation}
The first and second terms in Eqs.~(\ref{3dpvdos}) and~(\ref{2dpvdos}) define $g_1(\omega) \equiv f_1\left[ \omega; c_\alpha(\omega) \right]$ and $g_2(\omega) \equiv f_2\left[\omega; c_\alpha(\omega),\Gamma_\alpha(\omega) \right]$ in Eq.~(\ref{pvdos}), respectively.
Note that the second term $g_2(\omega)$ represents the nonphonon vDOS, $g_\text{nonphonon}(\omega)$ (see the discussion in Section~\ref{sectionnondeby}).

\begin{figure*}[t]
\centering
\includegraphics[width=0.95\textwidth]{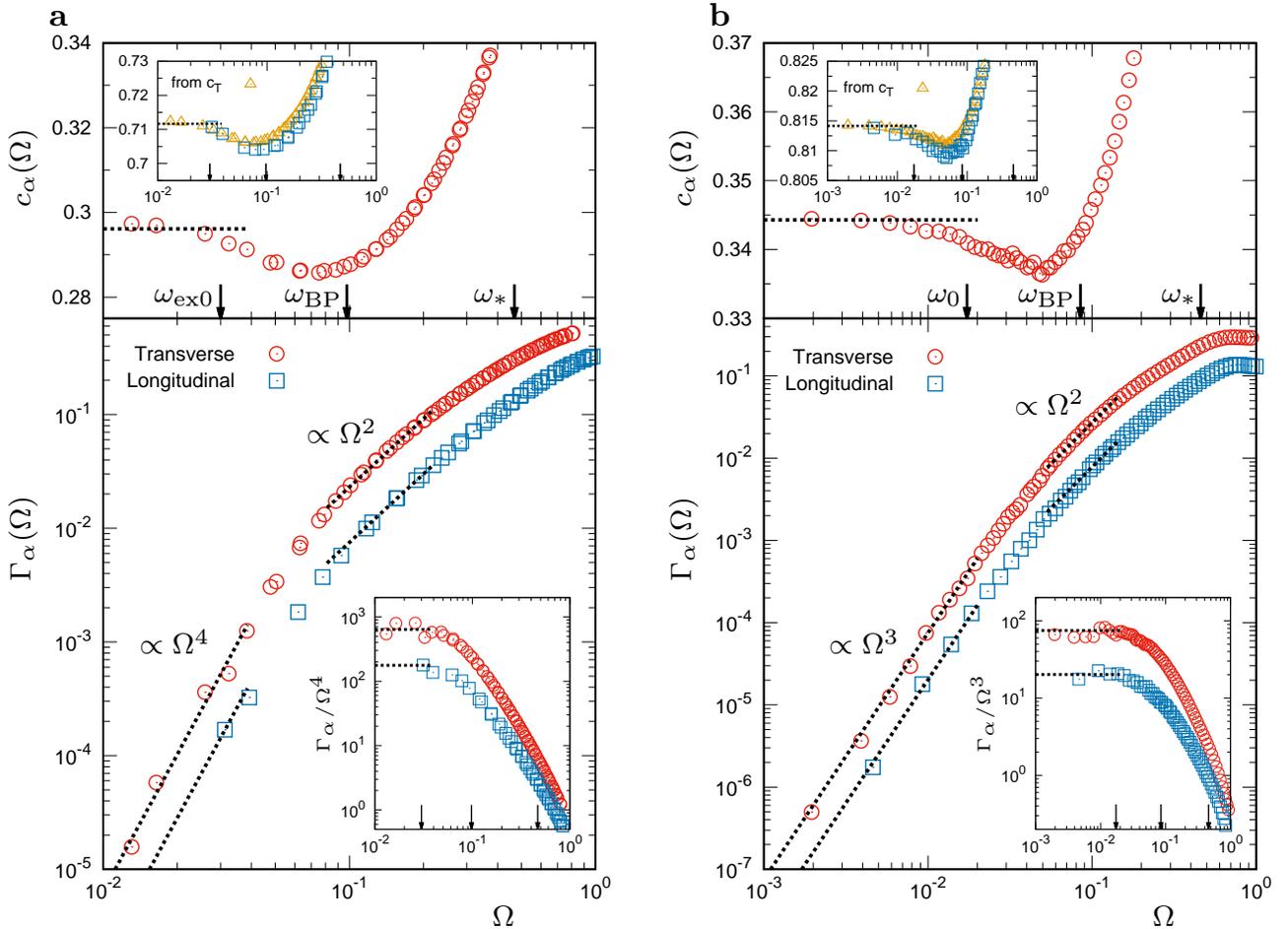}
\vspace*{0mm}
\caption{\label{fig1}
{Phonon transport in the amorphous solid models.}
Plots of the sound speed $c_\alpha (\Omega)$ and the attenuation rate $\Gamma_\alpha (\Omega)$ as functions of $\Omega$, for transverse ($\alpha =T$, red circles) and longitudinal ($\alpha = L$, blue squares) waves.
{\bf a} (left panels), The 3D model system ($d=3$).
{\bf b} (right panels), The 2D model system ($d=2$).
The packing pressure is $p=5 \times 10^{-2}$.
In the top panels, the characteristic frequencies of $\omega_\text{ex0}$~(3D), $\omega_0$~(2D), $\omega_\text{BP}$, and $\omega_\ast$ are indicated by arrows.
The insets in the top panels compare the longitudinal speed $c_L(\Omega)$ with the value predicted from the transverse speed $c_T(\Omega)$ as follows: $\sqrt{K/\rho + 4c_T(\Omega)^2/3}$ (3D) or $\sqrt{K/\rho + c_T(\Omega)^2}$ (2D).
The horizontal lines in the top panels indicate the macroscopic sound speeds, $c_{T0} = \sqrt{G/\rho}$ and $c_{L0} = \sqrt{(K + 4G/3)/\rho}$ (3D) or $c_{L0} = \sqrt{(K + G)/\rho}$ (2D).
The insets in the bottom panels show $\Gamma_\alpha(\Omega)/\Omega^{d+1}$ as a function of $\Omega$.
Both transverse and longitudinal waves exhibit behaviors of $\Gamma_{\alpha} \propto \Omega^{d+1}$ (Rayleigh scattering) at $\Omega \lesssim \omega_\text{ex0}$ (3D) or $\Omega \lesssim \omega_0$ (2D) and $\Gamma_{\alpha} \propto \Omega^2$ ($\Omega^2$ law) near $\Omega \sim \omega_\text{BP}$.
}
\end{figure*}

\section{Results and Discussion}
\subsection{Phonon transport properties}
We now study the phonon transport in the 3D amorphous solid. 
We excite a phonon with wavenumber $q$ at the initial time and analyze the decay profile of the velocity autocorrelation function to extract the propagation frequency $\Omega$ and the attenuation rate $\Gamma$ (see Section~\ref{methodphonon}).
The packing pressure is fixed at $p=5 \times 10^{-2}$ (the packing fraction is $\varphi \approx 0.73$).
Figure~\ref{fig1}a shows the obtained sound velocities $c_\alpha(\Omega) \equiv \Omega/q$ and the attenuation rates $\Gamma_\alpha(\Omega)$ for transverse ($\alpha = T$) and longitudinal ($\alpha = L$) waves.
Note that our measurement is in the zero-temperature harmonic limit.

We first focus on the BP regime, $\Omega \sim \omega_\text{BP}$.
In this regime, $c_\alpha$ shows a minimum value; i.e., the dispersion curve $\Omega(q)$ deviates from a straight line.
This is the so-called sound softening phenomenon, which is often observed in experiments~\cite{ruffle_2006,Masciovecchio_2006,Monaco_2009,Baldi_2010,Baldi_2011,Baldi_2016}.
$\Gamma_\alpha$ shows an $\Omega^2$ dependence, $\Gamma_\alpha \propto \Omega^2$.
We also find that the Ioffe-Regel (IR) limit for transverse waves is $\Omega_\text{TIR} \approx \omega_\text{BP}$~(see Appendix~\ref{IRsection}; see also Figs.~\ref{figs5} and~\ref{figs6a}).
These results indicate that the phonon does not propagate as a plane wave but rather exhibits dynamics characteristic of viscous damping~\cite{shintani_2008,Mizuno_2012}.
This can be understood in terms of the underlying vibrational eigenmodes as follows.
Because the eigenmodes at $\Omega \sim \omega_\text{BP}$ correspond to disordered and extended vibrations~\cite{Wyart_2005,Wyart_2006,Silbert_2005,Silbert_2009,Mizuno_2017} and the initially excited phonon is decomposed into these vibrational eigenmodes~\cite{taraskin_2000}, it immediately attenuates to become diffusive.
Refs.~\cite{Allen_1993,Feldman_1993} referred to this vibrational behavior as diffuson (nonpropagating, delocalized vibration).
Note that above the BP, $c_\alpha$ increases with increasing $\Omega$, which is referred to as sound hardening and was recently discussed in Ref.~\cite{Crespo_2016}.

As $\Omega$ decreases to the continuum limit, $\Omega \lesssim \omega_\text{ex0}$, we observe a clear crossover to Rayleigh scattering.
$c_\alpha$ converges to the macroscopic value predicted by the continuum mechanics: $c_{T0} = \sqrt{G/\rho}$ and $c_{L0} = \sqrt{(K + 4G/3)/\rho}$, where $\rho$ is the mass density and $K$ and $G$ are the bulk and shear moduli, respectively.
Note that we calculate $K$ and $G$ in the zero-temperature harmonic limit, by using the harmonic formulation described in Ref.~\cite{Mizuno3_2016}.
$\Gamma_\alpha$ shows an $\Omega^4$ dependence, $\Gamma_\alpha \propto \Omega^4$.
These observations reflect the Rayleigh scattering mechanism underlying the phonon propagation at $\Omega \lesssim \omega_\text{ex0}$.
This can again be understood in terms of the underlying vibrational eigenmodes as follows.
The eigenmodes at $\Omega \lesssim \omega_\text{ex0} $ consist of phonon modes and soft localized modes.
Because the initially excited phonon is decomposed mainly into the phonon modes
\footnote{
We measure the overlap $O_\text{loc}$ between the initially excited phonon $\dot{\mathbf{u}}^0_i$ and the soft localized modes $\mathbf{e}^k_i$ with participation ratio $P^k < 10^{-2}$~\cite{taraskin_2000}:
\begin{equation}
O_\text{loc} = \sum_{k; P^k < 10^{-2}} \sum_{i=1}^N \left| \dot{\mathbf{u}}^0_i \cdot \mathbf{e}^k_i \right|^2.
\end{equation}
We find that a few~$\%$ (at most) of the initially excited phonon is made up of soft localized modes.},
it attenuates slowly.

The crossover between $\Omega^2$ law and Rayleigh scattering has been observed in both experiments~\cite{ruffle_2006,Masciovecchio_2006,Monaco_2009,Baldi_2010,Baldi_2011,Baldi_2016} and numerical simulations~\cite{Monaco2_2009,Marruzzo_2013,Mizuno_2014}.
A recent work~\cite{Beltukov_2018} also found the crossover between diffusive regime and propagative regime by investigating the atomic response to a wave-packet excitation.
However, the precise location of the crossover frequency remains controversial.
The present work unambiguously links these two modes of phonon transport to the vibrational eigenmodes in the corresponding frequency regimes.
Therefore, we are able to identify the crossover frequency as the continuum limit frequency $\omega_\text{ex0}$, at which the nature of the underlying eigenmodes changes.
This result is consistent with the predictions of mean-field theories~\cite{schirmacher_2006,schirmacher_2007,Marruzzo_2013,Schirmacher_2015,Wyart_2010,DeGiuli_2014}.

We note that the IR limit for longitudinal waves, $\Omega_\text{LIR}$, is much higher than that for transverse waves~\cite{Wang_2015}: $\Omega_\text{LIR} \gg \Omega_\text{TIR} \approx \omega_\text{BP}$~(see Appendix~\ref{IRsection} and Figs.~\ref{figs5} and~\ref{figs6a}), so a longitudinal wave can propagate even at $\Omega > \omega_\text{BP}$.
This result indicates that although the disordered vibrational modes are dominant at $\Omega > \omega_\text{BP}$, the longitudinal phonon modes also exist in this frequency regime.
On the other hand, as pointed out by Refs.~\cite{Monaco2_2009,Marruzzo_2013}, longitudinal waves show similar transport properties as those of transverse waves.
Indeed, we determine the same crossover frequency $\omega_\text{ex0}$ for both transverse and longitudinal waves.
Heterogeneous elasticity theory~\cite{schirmacher_2006,schirmacher_2007,Marruzzo_2013,Schirmacher_2015} assumes that the shear modulus heterogeneity dominates compared with the bulk modulus heterogeneity, which is true in the present amorphous system~\cite{Mizuno2_2015}.
In this framework, the shear modulus heterogeneity induces the anomalous behaviors for both transverse and longitudinal waves
\footnote{
In this framework where the shear modulus heterogeneity dominates compared with the bulk modulus heterogeneity, the attenuation rate of transverse waves becomes larger than that of longitudinal waves, since the transverse waves are more sensitive to the shear modulus heterogeneity.
This is indeed consistent with our numerical observations.
}.
To check the validity of this scenario, we calculate the longitudinal sound speed as $\sqrt{K/\rho + 4c_T(\Omega)^2/3}$, in which the effects of the bulk modulus heterogeneity are neglected.
The inset at the top of Fig.~\ref{fig1}a shows that this value is close to the true longitudinal speed $c_L(\Omega)$ and thus supports this scenario.
However, Ref.~\cite{Mizuno_2014} reported that this property depends on the amorphous system and on the preparation procedures applied.

\begin{figure*}[t]
\centering
\includegraphics[width=0.95\textwidth]{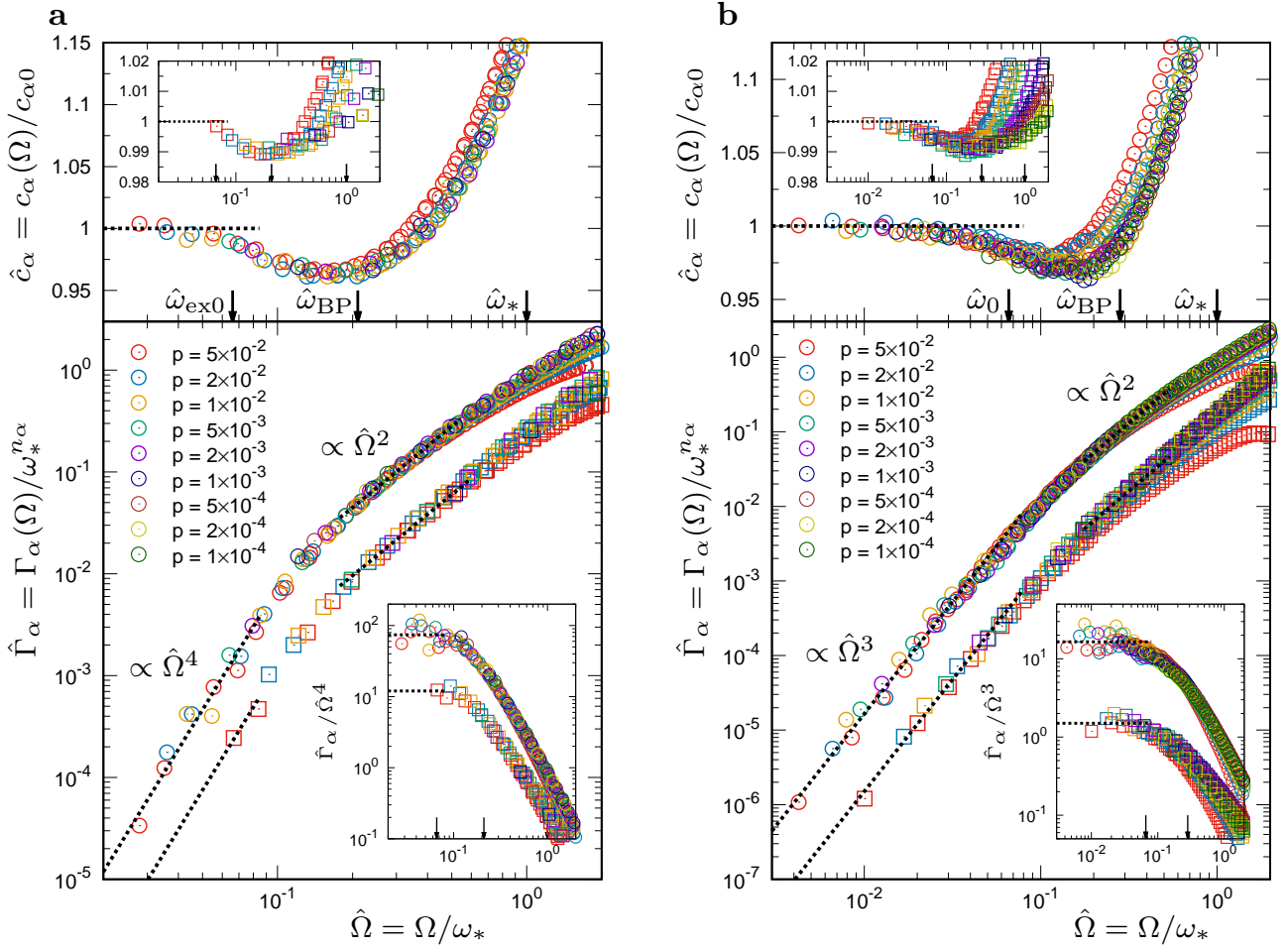}
\vspace*{0mm}
\caption{\label{fig2}
{Phonon transport at different pressures.}
Plots of the scaled sound speed $\hat{c}_\alpha = c_\alpha/c_{\alpha 0}$ and the attenuation rate $\hat{\Gamma}_\alpha = \Gamma_\alpha/\omega_\ast^{n_\alpha}$ as functions of the scaled frequency $\hat{\Omega} = \Omega /\omega_\ast$, for transverse ($\alpha =T$, circles) and longitudinal ($\alpha = L$, squares) waves.
{\bf a} (left panels), The 3D model system ($d=3$).
{\bf b} (right panels), The 2D model system ($d=2$).
The packing pressure ranges from $p=5 \times 10^{-2}$ to $1 \times 10^{-4}$.
In the top panels, the characteristic frequencies of $\hat{\omega}_\text{ex0} = \omega_\text{ex0}/\omega_\ast = 0.066$ (3D), $\hat{\omega}_0 = \omega_\text{0} / \omega_\ast = 0.066$ (2D), $\hat{\omega}_\text{BP} = \omega_\text{BP}/\omega_\ast = 0.21$ (3D), $\hat{\omega}_\text{BP} = 0.28$ (2D), and $\hat{\omega}_\ast  = \omega_\ast / \omega_\ast \equiv 1$ are indicated by arrows.
The horizontal lines in the top panels show the macroscopic values, $\hat{c}_\alpha = 1$.
The insets in the bottom panels show $\hat{\Gamma}_\alpha/\hat{\Omega}^{d+1}$ as a function of $\hat{\Omega}$.
The scaled $\hat{c}_\alpha$ and $\hat{\Gamma}_\alpha$ values collapse to a single curve at $\hat{\Omega} \lesssim \hat{\omega}_\text{BP}$ (and at $p \le 5\times 10^{-3}$ for 2D), with exponents of $n_T=1.0$ and $n_L=1.4$ (3D) or  $n_T=1.0$ and $n_L=1.6$ (2D).
For clarity, we multiply $\hat{\Gamma}_L$ by $0.5$ (3D) or $0.2$ (2D) in the plots.
See also Fig.~S1 of the Supplemental Material~\cite{supplement} for the collapse of $\hat{\Gamma}_T$.
}
\end{figure*}

\subsection{Packing pressure dependence}
We repeat the phonon transport calculations for various pressures over a wide range from $p = 5 \times 10^{-2}$ to $1 \times 10^{-4}$.
Below, we attempt to summarize the pressure dependence results in the form of scaling laws with $p$.
To this end, we introduce the scaled frequency $\hat{\Omega} \equiv \Omega/\omega_\ast$.
We recall that the three characteristic frequencies follow the same scaling law $\omega_\text{ex0} \propto \omega_\text{BP} \propto \omega_\ast \propto p^{1/2}$. 
More quantitatively, we can express the other two frequencies in scaled form as follows: $\hat{\omega}_\text{ex0} \equiv \omega_\text{ex0}/\omega_\ast = 0.066$ and $\hat{\omega}_\text{BP} \equiv \omega_\text{BP}/\omega_\ast = 0.21$~\cite{Mizuno_2017}. 
We explicitly plot these characteristic frequencies as functions of $p$ in Fig.~\ref{figs6a}a. 

The top panel of Fig.~\ref{fig2}a presents the scaled sound speed $\hat{c}_\alpha \equiv c_\alpha /c_{\alpha 0}$ as a function of $\hat{\Omega}$ for different pressures (where $c_{\alpha 0}$ is the macroscopic value). 
The results show a nice collapse to a universal curve at $\hat{\Omega} \lesssim \hat{\omega}_\text{BP}$; therefore, we obtain the scaling function for ${c}_\alpha$ in the following form: 
\begin{equation} \label{cscaling}
\begin{aligned}
c_\alpha(\Omega) = 
\left\{ \begin{aligned}
& c_{\alpha 0} f_{\alpha \text{min}} & (\Omega \sim \omega_\text{BP}), \\
& c_{\alpha 0} & (\Omega \lesssim \omega_\text{ex0}),
\end{aligned} \right.
\end{aligned} 
\end{equation}
where $f_{T \text{min}} = 0.96$ and $f_{L \text{min}} = 0.99$ are constants that measure the reduction in ${c}_\alpha$ near the BP. 
We also plot the scaled attenuation rate $\hat{\Gamma}_\alpha \equiv \Gamma_\alpha (\Omega) /\omega_\ast^{n_\alpha}$ in the bottom panel of Fig.~\ref{fig2}a. 
Here, we use the exponent $n_\alpha$ as a fitting parameter and find that $\hat{\Gamma}_\alpha$ nicely collapses to a universal curve with $n_T = 1.0$ (transverse waves) and $n_L = 1.4$ (longitudinal waves).
(Figure~S1a of the Supplemental Material~\cite{supplement} plots $(\Gamma_T/ \omega_\ast^{n_T}) / \hat{\Omega}^4$ versus $\hat{\Omega}$ for several different values of $n_T$, from $n_T = 1.5$ to $0.5$, where we see the best collapse with $n_T = 1.0$.)
This suggests the following form for the scaling function of $\hat{\Gamma}_\alpha$: 
\begin{equation} \label{gammascaling}
\begin{aligned}
\Gamma_\alpha(\Omega) = 
\left\{ \begin{aligned}
& B_{\alpha \text{BP}} \omega_\ast^{n_\alpha} \left( \frac{\Omega}{\omega_\ast} \right)^2 & (\Omega \sim \omega_\text{BP}), \\
& B_{\alpha 0} \omega_\ast^{n_\alpha} \left( \frac{\Omega}{\omega_\ast} \right)^4 & (\Omega \lesssim \omega_\text{ex0}),
\end{aligned} \right. 
\end{aligned}
\end{equation}
where $B_{T \text{BP}} = 1.1$, $B_{L \text{BP}}=0.48$, $B_{T 0} = 74$, and $B_{L 0}=24$ are constants that represent the strength of attenuation.

Effective medium theory~\cite{Wyart_2010,DeGiuli_2014} predicts scaling laws with $p$ or $\omega_\ast \propto p^{1/2}$.
We therefore test the validity of this theory by comparing the scaling laws indicated by our numerical results with the theoretical predictions.
To directly compare our results with theoretical predictions of Ref.~\cite{DeGiuli_2014}, we plot the scattering length $\ell_{\alpha} \equiv 2c_\alpha / \Gamma_\alpha$ in Fig.~S2a of the Supplemental Material~\cite{supplement}.
Our results yield the following scaling laws for transverse waves:
\begin{equation} \label{lscaling}
\begin{aligned}
\ell_T(\Omega) & \approx
\left\{ \begin{aligned}
& \frac{ 2C_0 }{ B_{T \text{BP}} } \frac{ \omega_\ast^{3/2} } {\Omega^{2}} \propto \frac{ \omega_\ast^{3/2} } {\Omega^{2}} & (\Omega \sim \omega_\text{BP}), \\
& \frac{ 2C_0 }{ B_{T 0} } \frac{ \omega_\ast^{7/2} }{\Omega^{4}} \propto  \frac{ \omega_\ast^{7/2} }{\Omega^{4}} & (\Omega \lesssim \omega_\text{ex0}),
\end{aligned} \right.
\end{aligned}
\end{equation}
where we use $c_{T} \approx c_{T0} \propto \sqrt{G} \propto \omega_\ast^{1/2}$, which is numerically confirmed to correspond to $c_{T0} = C_0 \omega_\ast^{1/2}$ with $C_0 = 0.47$~\cite{Mizuno3_2016}.
In the BP regime, our scaling result is consistent with the theoretical prediction~\cite{DeGiuli_2014}.
However, in the Rayleigh scattering regime, the theory predicts $\ell_T(\Omega) \propto \omega_\ast^{3} / \Omega^{4}$~\cite{DeGiuli_2014}, which is different from our result by a factor of $\omega_\ast^{-1/2}$.  
This means that the theory significantly overestimates the scattering length (underestimates the strength of the phonon scattering) and the difference between the theory and numerical result even diverges at the jamming transition.

\begin{figure*}[t]
\centering
\includegraphics[width=0.95\textwidth]{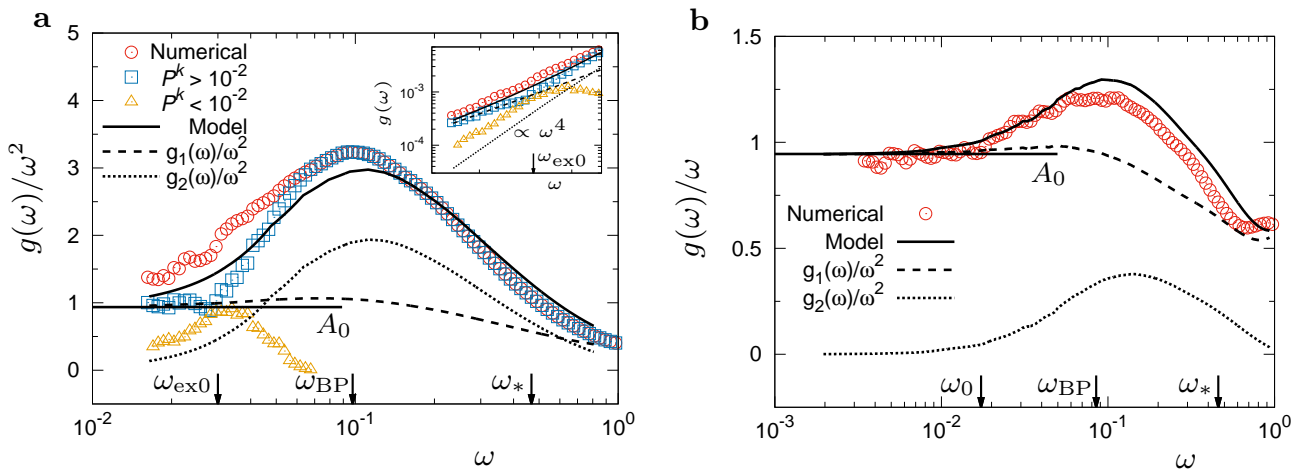}
\vspace*{0mm}
\caption{\label{fig3}
{Comparison between the numerical vDOS values obtained through vibrational mode analysis and the vDOS of the generalized Debye model.}
Plots of the reduced vDOS $g(\omega)/\omega^{d-1}$ as a function of $\omega$.
{\bf a}, The 3D model system ($d=3$).
{\bf b}, The 2D model system ($d=2$).
The packing pressure is $p=5 \times 10^{-2}$.
The inset in {\bf a} shows $g(\omega)$.
The symbols represent the numerical values from the vibrational mode analysis reported in Ref.~\cite{Mizuno_2017}.
For the 3D system in {\bf a}, the vDOS of the extended modes, $g_\text{ex}(\omega)$ for the participation ratio $P^k > 10^{-2}$, and that of the localized modes, $g_\text{loc}(\omega)$ for $P^k < 10^{-2}$, are also plotted.
The lines represent the vDOS of the generalized Debye model as given in Eq.~(\ref{pvdos}) [Eq.~(\ref{3dpvdos}) for the 3D and Eq.~(\ref{2dpvdos}) for the 2D].
The vDOS in Eq.~(\ref{pvdos}) is composed of two terms, one related to the dispersion curve ($g_1(\omega) \equiv f_1 \left[ \omega; c_\alpha(\omega) \right]$) and the other related to the sound broadening ($g_2(\omega) \equiv f_2 \left[ \omega; c_\alpha(\omega),\Gamma_\alpha(\omega) \right]$).
The horizontal lines indicate the Debye value $A_0$ as calculated from the macroscopic elastic moduli.
}
\end{figure*}

\subsection{Generalized Debye model}
We next apply the generalized Debye model to calculate the vDOS of the vibrational eigenmodes~\cite{schirmacher_2006,schirmacher_2007,Marruzzo_2013,Schirmacher_2015}.
We will show that the generalized Debye model well reproduces the vDOS from the data of $c_\alpha$ and $\Gamma_\alpha$, and therefore, it relates the phonon transport properties to the vDOS of the underlying vibrational modes.

As already mentioned, the IR limit for transverse waves, $\Omega_\text{TIR}$, is located near the BP.
The phonon approximation, $\Omega \gg \Gamma_\alpha$, is therefore valid below the BP, at $\Omega \lesssim \omega_\text{BP}$.
In this approximation, the generalized Debye model gives the Green function $\tilde{G}_\alpha(q,\omega)$ in terms of $c_\alpha(\omega)$ and $\Gamma_\alpha(\omega)$, as shown in Eq.~(\ref{greenf3}).
The vDOS $g(\omega)$ is then formulated as shown in Eq.~(\ref{3dpvdos}) for the 3D case:
\begin{equation} \label{pvdos}
\begin{aligned}
g(\omega) & = g_1(\omega) + g_2(\omega), \\
& \equiv f_1 \left[ \omega; c_\alpha(\omega) \right] + f_2 \left[ \omega; c_\alpha(\omega),\Gamma_\alpha(\omega) \right],
\end{aligned}
\end{equation}
where $g_1(\omega) \equiv f_1 \left[ \omega; c_\alpha(\omega) \right]$ is a functional of $c_\alpha(\omega)$, whereas $g_2(\omega) \equiv f_2 \left[ \omega; c_\alpha(\omega),\Gamma_\alpha(\omega) \right]$ is a functional of both $c_\alpha(\omega)$ and $\Gamma_\alpha(\omega)$.
Please refer to Section~\ref{sectionGDM} and Eq.~(\ref{3dpvdos}) for the 3D and Eq.~(\ref{2dpvdos}) for the 2D, for detail of formulations.

Figure~\ref{fig3}a compares the reduced vDOS $g(\omega)/\omega^2$ calculated with Eq.~(\ref{pvdos}) and the numerical values reported in Ref.~\cite{Mizuno_2017} at $p=5 \times 10^{-2}$.
Note that these numerical values were obtained through vibrational mode analysis, i.e., by considering the statistics of the vibrational eigenmodes.
The generalized Debye model overall captures the numerical values of $g(\omega)$.
(The quantitative difference from the numerical values might arise from the phonon approximation that is adopted when formulating Eq.~(\ref{pvdos}) (Eq.~(\ref{3dpvdos})) in the generalized Debye model.)
In particular, Eq.~(\ref{pvdos}) captures the BP at $\omega = \omega_\text{BP}$.
Previous works~\cite{Monaco2_2009,Monaco_2009} have argued that the BP originates from the deformation of the dispersion curve, i.e., from the first term $g_1(\omega) \equiv f_1 \left[ \omega; c_\alpha(\omega) \right]$.
However, our results show that $g_1(\omega)/\omega^2$ exhibits only a tiny peak near $\omega_\text{BP}$, which cannot explain the BP.
To express the BP correctly, we need to consider the second term $g_2(\omega) \equiv f_2 \left[ \omega; c_\alpha(\omega),\Gamma_\alpha(\omega) \right]$, which arises from the sound broadening $\Gamma_\alpha$.
We note that although the phonon approximation is valid at low frequencies, $\omega \lesssim \omega_\text{BP}$, Eq.~(\ref{pvdos}) seems to reproduce the numerical values even at higher frequencies, $\omega_\text{BP} < \omega \lesssim \omega_\ast$.

Eq.~(\ref{pvdos}) captures the numerical values in the continuum limit $\omega \lesssim \omega_\text{ex0}$, where the vibrational modes are composed of phonon modes and soft localized modes.
$g_1(\omega)/\omega^2$ converges to the Debye level $A_0$, whereas $g_2(\omega)/\omega^2$, which represents the nonphonon contribution, converges only slowly to zero.
$g_2(\omega)$ originates from $\Gamma_\alpha$ or, more specifically, $\Gamma_T$ for transverse waves, which is much larger than its longitudinal counterpart $\Gamma_L$.
Because the nonphonon modes in this frequency regime are detected as soft localized modes~\cite{Mizuno_2017,Lerner_2016}, $g_2(\omega)$ should correspond to the vDOS of the soft localized modes.
Indeed, the inset of Fig.~\ref{fig3}a and Eq.~(\ref{3dpvdos}) (in the low-$\omega$ limit) indicate that $g_2(\omega) \propto \Gamma_T \propto \omega^4$, which fully coincides with the $\omega^4$ law for soft localized modes.

\begin{figure}[t]
\centering
\includegraphics[width=0.475\textwidth]{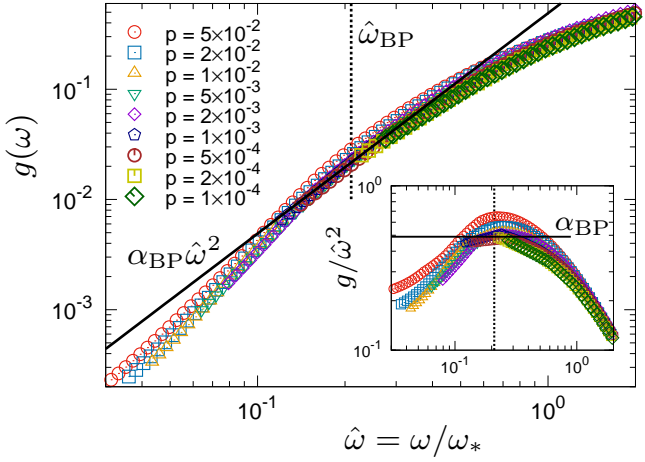}
\vspace*{0mm}
\caption{\label{fig4}
{The vDOSs of the generalized Debye model at different pressures in the 3D model system.}
$g(\omega)$ is plotted against the scaled frequency $\hat{\omega}=\omega/\omega_\ast$.
The inset is the same as the main panel but for the reduced vDOSs $g/\hat{\omega}^2$.
At low pressures and near $\hat{\omega}_\text{BP} = 0.21$, the vDOSs collapse to the following non-Debye scaling law: $g(\omega) = \alpha_\text{BP} \hat{\omega}^2$ with $\alpha_\text{BP} = 0.49$.
}
\end{figure}

\begin{figure}[t]
\centering
\includegraphics[width=0.475\textwidth]{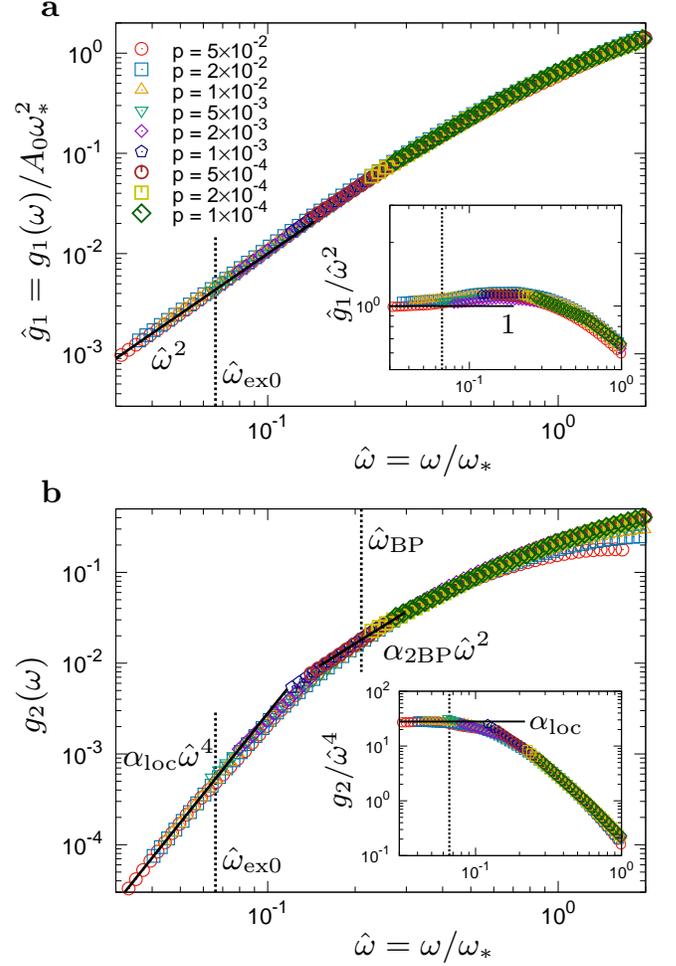}
\vspace*{0mm}
\caption{\label{fig5}
{The vDOSs of the generalized Debye model, $g_1(\omega)$ and $g_2(\omega)$, in the 3D model system.}
{\bf a}, Plot of the scaled vDOS $\hat{g}_1 = g_1(\omega)/A_0 \omega_\ast^2$ against the scaled frequency $\hat{\omega}=\omega/\omega_\ast$.
{\bf b}, Plot of ${g}_2(\omega)$ against $\hat{\omega}$.
The insets are the same as the main panels but for the reduced vDOSs, $\hat{g}_1/\hat{\omega}^2$ in {\bf a} and ${g}_2/\hat{\omega}^4$ in {\bf b}.
Below $\hat{\omega}_\text{ex0}=0.066$, the vDOSs collapse to $g_1(\omega) = A_0 \omega^2$ and $g_2(\omega) = \alpha_\text{loc} (\omega/\omega_\ast)^4$ with $\alpha_\text{loc} = 28$.
In addition, near $\hat{\omega}_\text{BP}=0.21$, $g_2(\omega) = \alpha_\text{2BP} (\omega/\omega_\ast)^2$ with $\alpha_\text{2BP} = 0.41$.
}
\end{figure}

\subsection{Phonon scattering and non-Debye scaling law}~\label{sectionnondeby}
We now extend the analysis of the vDOS of the generalized Debye model to different pressures $p$ and establish the scaling behaviors of the vDOS.
We will show that the non-Debye scaling laws for the vDOS are intimately related to the phonon scattering laws for $\Omega^2$ law, $\Gamma_\alpha = B_{\alpha \text{BP}} \omega_\ast^{n_\alpha} \left( {\Omega}/{\omega_\ast} \right)^2$, and Rayleigh scattering, $\Gamma_\alpha = B_{\alpha 0} \omega_\ast^{n_\alpha} \left( {\Omega}/{\omega_\ast} \right)^4$. 

Figure~\ref{fig4} plots $g(\omega)$ and the reduced $g(\omega)/\hat{\omega}^2$ as functions of the scaled $\hat{\omega} \equiv \omega/\omega_\ast$ for various $p$.
Near the BP, at $\hat{\omega}_\text{BP} = 0.21$, the vDOS follows a non-Debye scaling law, $g(\omega) = \alpha_\text{BP} (\omega/\omega_\ast)^2$ with $\alpha_\text{BP}=0.49$.
In addition, Figure~\ref{fig5} plots the scaled $\hat{g}_1 \equiv g_1(\omega)/A_0 \omega_\ast^2$ and $g_2(\omega)$ separately.
We observe that $\hat{g}_1(\omega)$ and $g_2(\omega)$ collapse to universal functions of $\hat{\omega}$ for different pressures.
In the continuum limit $\hat{\omega} \lesssim \hat{\omega}_\text{ex0} = 0.066$, $g_1(\omega)$ converges to the Debye vDOS, $g_1(\omega) = A_0 \omega^2$, whereas $g_2(\omega)$, which corresponds to the vDOS of the soft localized modes, converges to another non-Debye scaling law, $g_2(\omega) = \alpha_\text{loc} (\omega/\omega_\ast)^4$ with $\alpha_\text{loc} = 28$.
At higher frequencies, $g_2(\omega)$ becomes much larger than $g_1(\omega)$, and the total vDOS becomes $g(\omega) \approx g_2(\omega)$.
At $\omega \sim {\omega}_\text{BP}$ in particular, $g_2(\omega) = \alpha_\text{2BP} (\omega/\omega_\ast)^2$ with $\alpha_\text{2BP}=0.41$, which confirms that $g_2(\omega) \approx g(\omega) = \alpha_\text{BP} (\omega/\omega_\ast)^2$ ($\alpha_\text{2BP} = 0.41$ is close to $\alpha_\text{BP} = 0.49$).
We can therefore describe the nonphonon vDOS as $g_\text{nonphonon}(\omega) \approx g_2(\omega) = G(\omega/\omega_\ast)$, where $G(\hat{\omega}) = \alpha_\text{BP} \hat{\omega}^2$ for $\hat{\omega} \sim \hat{\omega}_\text{BP} = 0.21$ and $G(\hat{\omega}) = \alpha_\text{loc} \hat{\omega}^4$ for $\hat{\omega} \lesssim \hat{\omega}_\text{ex0} = 0.066$.
All of these results are completely consistent with the results of the vibrational mode analysis in Eq.~(\ref{vdos})~\cite{Mizuno_2017}.

These results enable us to directly relate the attenuation rate $\Gamma_\alpha$ in Eq.~(\ref{gammascaling}) to the nonphonon part of the vDOS $g_\text{nonphonon}(\omega)$ in Eq.~(\ref{vdos}) as follows.
Since $c_T < c_L$ and $\Gamma_T \gg \Gamma_L$, we neglect the contribution from longitudinal waves in Eq.~(\ref{pvdos}) (Eq.~(\ref{3dpvdos})).
We can then express $g_\text{nonphonon}(\omega) \approx g_2(\omega)$ as
\begin{equation}
\begin{aligned} \label{vdosgamma}
& g_\text{nonphonon}(\omega) \approx \frac{4}{ \pi q_D^2} \frac{\Gamma_T (\Omega = \omega)}{c_{T0}^2}, \\
& \quad  =
\left\{ \begin{aligned}
& \frac{4 B_{T\text{BP}} } { \pi q_D^2 C_0^2} \left( \frac{\omega}{\omega_\ast} \right)^2 \propto \left( \frac{\omega}{\omega_\ast} \right)^2 & (\omega \sim \omega_\text{BP}), \\
& \frac{4 B_{T 0}}{ \pi q_D^2 C_0^2} \left( \frac{\omega}{\omega_\ast} \right)^4 \propto \left( \frac{\omega}{\omega_\ast} \right)^4 & (\omega \lesssim \omega_\text{ex0}).
\end{aligned} \right.
\end{aligned}
\end{equation}
This expression suggests that there are relationships between the parameters in the scaling functions for the phonon transport modes and those of the vDOS: ${4 B_{T\text{BP}} } / ({\pi q_D^2 C_0^2}) \sim \alpha_\text{BP}$ and ${4 B_{T 0}} / ({\pi q_D^2 C_0^2}) \sim \alpha_\text{loc}$. 
We confirm that the values of ${4 B_{T\text{BP}} } / ({\pi q_D^2 C_0^2}) = 0.38$ and ${4 B_{T 0}} / ({\pi q_D^2 C_0^2}) = 25$ are indeed comparable to those of $\alpha_\text{BP}=0.66$ and $\alpha_\text{loc} = 58$ as evaluated through vibrational mode analysis~\cite{Mizuno_2017}.
(We cannot expect that values of ${4 B_{T\text{BP}} } / ({\pi q_D^2 C_0^2})$ and ${4 B_{T 0}} / ({\pi q_D^2 C_0^2})$, obtained from the generalized Debyel model that assumes the phonon approximation, exactly coincide with those of $\alpha_\text{BP}$ and $\alpha_\text{loc}$, obtained from the statistics of the vibrational eigenmodes, respectively.)
Eq.~(\ref{vdosgamma}) demonstrates that the term $\Gamma_T / c_{T0}^{2} \approx 2 / (\ell_T c_{T0})$ determines both the $\omega$ and $p$ dependences of $g_\text{nonphonon}(\omega)$.

We therefore conclude that the attenuation rate and the nonphonon vDOS are intimately related.
In the continuum limit $\omega \lesssim \omega_\text{ex0}$, the existence of the soft localized modes enhances the attenuation rate, which, in turn, results in the excess value of the vDOS represented by the term $g_2(\omega)$.
Notably, mean-field theories~\cite{schirmacher_2006,schirmacher_2007,Marruzzo_2013,Schirmacher_2015,Wyart_2010,DeGiuli_2014} neglect the vDOS of the soft localized modes and concomitantly underestimate the Rayleigh scattering amplitude.
In particular, the theory of~\cite{Wyart_2010,DeGiuli_2014} predicts $g_\text{nonphonon}(\omega) \propto (\ell_T c_{T0})^{-1} \propto \omega^4 / \omega_\ast^{7/2}$ at $\omega \lesssim \omega_\text{ex0}$, which differs from our result by a factor of $\omega_\ast^{1/2}$.

\subsection{The unstressed system}~\label{secunstress}
The unstressed system is defined as the system with all of the particles' contacts replaced with relaxed springs (see Section~\ref{system}).
Previous works~\cite{Mizuno_2017,Lerner2_2017} have shown that in the unstressed system, the soft localized modes are strongly suppressed, and the vDOS rapidly converges to the Debye level.
Therefore, one may expect that the phonon scattering in the unstressed system will be much weaker than that in the original stressed system.
The results of the unstressed system are presented in Figs.~S3 to S13 of the Supplemental Material~\cite{supplement}.

We first calculate the sound speed $c_\alpha(\Omega)$ and the attenuation rate $\Gamma_\alpha(\Omega)$ of the unstressed version of the configuration at $p=5 \times 10^{-2}$.
We observe the Rayleigh scattering law, $\Gamma_\alpha \propto \Omega^4$, even in the unstressed system (see Fig.~S3a).
However, as quantitatively estimated below, the coefficient of the Rayleigh scattering law is much smaller than in the original system.
This result supports the intimate connection between the strong attenuation and the presence of soft localized modes in the original stressed system.

We next analyze the pressure dependences of $c_\alpha$ and $\Gamma_\alpha$ (see Fig.~S4a).
We obtain the same pressure dependences of $c_\alpha$, $\Gamma_\alpha$, and $\ell_\alpha$ given in Eqs.~(\ref{cscaling}),~(\ref{gammascaling}), and (\ref{lscaling}), respectively. 
In the Rayleigh scattering regime $\omega < \omega_0$ in particular, $\Gamma_\alpha$ follows the scaling law $\Gamma_\alpha = B_{\alpha 0} \omega_\ast^{n_\alpha} \left({\Omega}/{\omega_\ast} \right)^4$ ($n_T=1.0$ and $n_L=1.4$, see also Fig.~S5a) even in the unstressed system.
The only difference between the stressed and unstressed systems lies in the magnitudes of the prefactors: $B_{T 0}=0.13$ and $B_{L 0}=0.10$ in the unstressed system, which are two orders of magnitude smaller than the values of $B_{T 0} = 74$ and $B_{L 0} = 24$ in the stressed system.
This result is somewhat surprising and suggests that effective medium theory~\cite{Wyart_2010,DeGiuli_2014} does \textit{not} correctly describe the phonon scattering even in the unstressed system.

We also apply the generalized Debye model to calculate the vDOS of the unstressed system (see Figs.~S6a and~S7).
The reduced nonphonon vDOS $g_2(\omega)/\omega^2$ decreases rapidly with decreasing $\omega$ and becomes negligible in the continuum limit, $\omega \lesssim \omega_\text{0}$.
Accordingly, the total $g(\omega)/\omega^2$ converges to the Debye level $A_0$.
This result is consistent with the numerical values found through vibrational mode analysis~\cite{Mizuno_2017}.
However, we remark that this rapid convergence occurs simply because the prefactors of $B_{T 0}$ and $B_{L 0}$ are very small.
Because the scaling behaviors of the phonon transport modes are unchanged between the stressed and unstressed systems, the generalized Debye model predicts the same scaling behavior of the nonphonon vDOS, $g_\text{nonphonon}(\omega) \approx g_2(\omega) \propto \left( {\omega}/{\omega_\ast} \right)^4$, even in the unstressed system (see Fig.~S9a).

\subsection{Length scales in the amorphous solid}
In this section, we will discuss our observations with regard to the characteristic length scales in the amorphous solid.
We first look at the continuum limit of phonon transport, which is measured with respect to the wavelength at $\Omega = \omega_{\text{ex0}}$, $\lambda_{\alpha 0} \equiv 2\pi c_{\alpha 0}/\omega_{\text{ex0}}$:
\begin{equation} \label{length3}
\lambda_{T0} \propto \omega_\ast^{-1/2} \propto p^{-1/4}, \qquad \lambda_{L0} \propto \omega_\ast^{-1} \propto p^{-1/2}.
\end{equation}
Early work~\cite{Silbert_2005} studied the wavelengths at $\Omega = \omega_\ast$, which follow the same scaling laws as those of $\lambda_{T0}$ and $\lambda_{L0}$.
$\lambda_{T0}$ and $\lambda_{L0}$ are naturally expected to correspond to the continuum limit of elastic response that has been studied in previous works~\cite{Ellenbroek_2006,Lerner_2014,Karimi_2015,Baumgarten_2017}.
In particular, Ref.~\cite{Karimi_2015} observed that the elastic response to a local force converges to the continuum limit (as predicted by continuum mechanics) at the length scales of $\xi_T \propto p^{-0.25}$ and $\xi_L \propto p^{-0.4}$ for the transverse and longitudinal components, respectively.
The scaling behaviors of $\xi_T$ and $\xi_L$ are consistent with those of $\lambda_{T0}$ and $\lambda_{L0}$, respectively.

For the transverse length scale $\xi_T$, Ref.~\cite{Lerner_2014} claimed that the elastic response is dominated by anomalous vibrational modes at $\omega_\ast$.
Effective medium theory~\cite{Wyart_2010,DeGiuli_2014} and, more recently, the variational argument~\cite{Yan_2016} predict that the spatial correlation of the particle displacements in these modes should extend over the length scale $\ell_c \propto p^{-1/4}$.
It has therefore been argued that $\xi_T$ is related to $\ell_c$, as $\xi_T \propto \ell_c \propto p^{-1/4}$~\cite{Lerner_2014}.
$\ell_c$ is also related to the scattering length at $\omega_\text{BP}$, as $\ell_T(\Omega = \omega_\text{BP}) \propto \omega_\ast^{-1/2} \propto p^{-1/4}$~\cite{Xu_2009,Vitelli_2010}.
In addition, we recently observed the length scale $\ell_c$ by measuring the size of a localized region in the soft localized modes~\cite{Shimada_2018}.

Furthermore, we study two additional length scales, $D_T$ and $D_L$, for transverse and longitudinal waves.
At $\Omega \lesssim \omega_\text{ex0}$, we can define a length $D$ that characterizes the structural disorder responsible for the Rayleigh scattering~\cite{Vitelli2_2010}.
To do this, we solve the scattering problem for elastic waves~\cite{Bhatia_1967}.
Let us consider a situation in which an elastic wave propagates in an elastic medium with scattering sources.
If we assume that its wavelength is much longer than the length $D$ of the scattering sources, we can derive the Rayleigh scattering law and formulate the attenuation rate as follows~\cite{Vitelli2_2010,Bhatia_1967,West_1984} (see Eq.~(\ref{waveeq9}) in Appendix~\ref{scattering}):
\begin{equation} \label{length1}
\Gamma_\alpha = \frac{\delta \gamma_\alpha^2}{4 \pi} \left( \frac{D_\alpha}{c_{\alpha 0}} \right)^3 \Omega^4 \propto \left( \frac{D_\alpha}{c_{\alpha 0}} \right)^3 \Omega^4,
\end{equation}
where $\delta \gamma_\alpha$ represents the strength of the elastic inhomogeneity (the scattering sources): $\delta \gamma_T = \delta G / G$ and $\delta \gamma_L = \delta (K + 4G/3) / (K + 4G/3)$ are for transverse and longitudinal waves, respectively (see Appendix~\ref{scattering}).
We remark that transverse and longitudinal waves are scattered by different elastic inhomogeneities, namely, shear ($G$) and longitudinal ($K+4G/3$) moduli inhomogeneities, respectively
\footnote{
$D_\alpha$ and $\delta \gamma_\alpha$ are not independent quantities.
$\delta \gamma_\alpha$ measures the strength of the elastic inhomogeneities for the length $D_\alpha$, and $D_\alpha$ measures the length scale of the elastic inhomogeneities for the strength $\delta \gamma_\alpha$.
In the present work, we fix the value of $\delta \gamma_\alpha$ and measure the pressure dependence of $D_\alpha$.
}.
Hence, the lengths $D_T$ and $D_L$ of these scattering sources are different.
We also note that for the case of a polycrystalline solid, $D$ is given by the typical grain size~\cite{Bhatia_1967,West_1984}; however, in an amorphous solid, $D$ cannot be directly observed from the static structure but instead can be determined in an indirect manner from the elastic response and the phonon transport.

By comparing Eq.~(\ref{length1}) with our simulation results as given in Eq.~(\ref{gammascaling}), we obtain
\begin{equation} \label{length2}
D_T \propto \omega_\ast^{-0.5} \propto p^{-0.25}, \qquad D_L \propto \omega_\ast^{-0.87} \propto p^{-0.43}.
\end{equation}
These scaling laws for $D_T$ and $D_L$ are consistent with those for $\lambda_T$ and $\lambda_L$ in Eq.~(\ref{length3}) and with those for $\xi_T$ and $\xi_L$~\cite{Ellenbroek_2006,Lerner_2014,Karimi_2015,Baumgarten_2017}.
This coincidence suggests that the size of the Rayleigh scattering sources is related to the length scale of the continuum limits of phonon transport and elastic response.
For transverse waves, $\lambda_T$, $\xi_T$, and $D_T$ might be controlled by the correlation length $\ell_c$ of the vibrational modes at $\omega_\ast$~\cite{Lerner_2014,Yan_2016}.
In the previous Section~\ref{secunstress}, the phonon transport modes were shown to be the same in the stressed and unstressed systems apart from the large difference in their prefactors.
This result is consistent with the discussion in this section, because the scaling behavior of $\ell_c$ is similarly known to be the same between the stressed and unstressed systems apart from a large difference in the prefactors~\cite{Lerner_2014,DeGiuli_2014}.

\begin{figure}[t]
\centering
\includegraphics[width=0.475\textwidth]{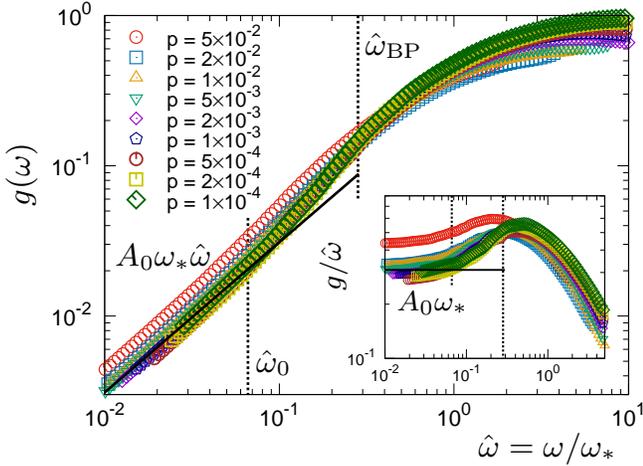}
\vspace*{0mm}
\caption{\label{fig6}
{The vDOSs of the generalized Debye model at different pressures in the 2D model system.}
$g(\omega)$ is plotted against the scaled frequency $\hat{\omega}=\omega / \omega_\ast$.
The inset is the same as the main panel but for the reduced vDOSs $g/\hat{\omega}$.
At low pressures, the vDOSs collapse to a universal function of the scaled frequency $\hat{\omega}$.
Below $\hat{\omega}_\text{0}=0.066$, $g(\omega)$ converges to the Debye vDOS, $g(\omega) = A_0 \omega$.
}
\end{figure}

\begin{figure}[t]
\centering
\includegraphics[width=0.475\textwidth]{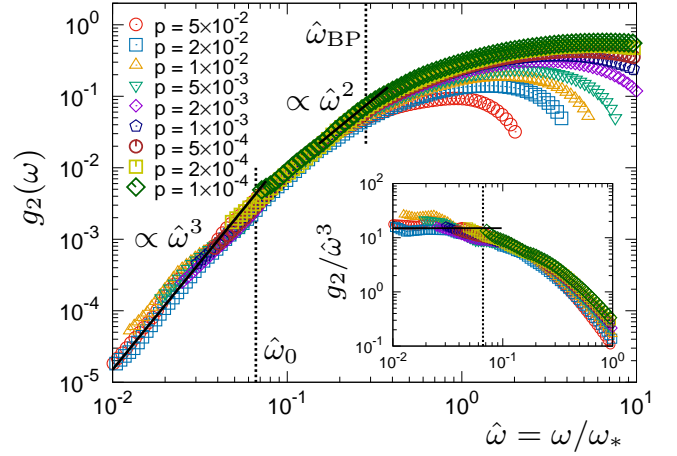}
\vspace*{0mm}
\caption{\label{figs10}
{The nonphonon vDOSs of the generalized Debye model in the 2D model system.}
The vDOSs $g_2(\omega)$ are plotted against the scaled frequency $\hat{\omega}=\omega / \omega_\ast$ for different pressures.
Around $\hat{\omega}_\text{BP}=0.28$, $g_2(\omega) \propto (\omega/\omega_\ast)^2$ is observed.
Below $\hat{\omega}_\text{0}=0.066$, the vDOSs collapse to $g_2(\omega) \propto (\omega/\omega_\ast)^3$.
}
\end{figure}

\begin{figure}[t]
\centering
\includegraphics[width=0.475\textwidth]{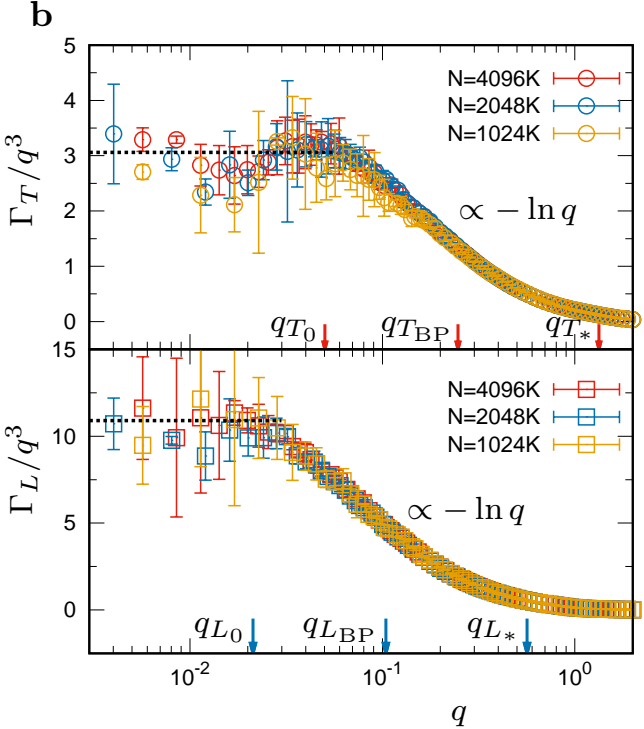}
\vspace*{0mm}
\caption{\label{figs23}
{Wavenumber dependence of the sound attenuation rate for the 2D model system.}
Plots of $\Gamma_\alpha/q^{3}$ as a function of $q$, for transverse ($\alpha =T$, circles) and longitudinal ($\alpha = L$, squares) waves.
The packing pressure is $p=5 \times 10^{-2}$.
Data are plotted for different system sizes of $N=1024000$, $2048000$, $4096000$.
In the panels, the characteristic wavenumbers of $q_{\alpha_\text{0}} \equiv \omega_\text{0}/c_{\alpha 0}$, $q_{\alpha_\text{BP}} \equiv \omega_\text{BP}/c_{\alpha 0}$, and $q_{\alpha_\ast} \equiv \omega_\ast/c_{\alpha 0}$ are indicated by arrows.
The horizontal line shows the value of $\Gamma_\alpha/q^3 = B_{\alpha 0} \omega_\ast^{n_\alpha-3} c_{\alpha 0}^3$, where $B_{\alpha 0} \omega_\ast^{n_\alpha-3}$ is the coefficient of $\Gamma_\alpha$ in the Rayleigh scattering regime, $\Gamma_\alpha = B_{\alpha 0} \omega_\ast^{{n_\alpha}} (\Omega/\omega_\ast)^{3}$ (see Eq.~(\ref{gammascaling2d})).
The error bars are estimated by using the values from different propagating directions of $x$ and $y$ and different initial phases of $\psi = 0$ and $\pi/2$.
A logarithmic scaling of $\Gamma_{\alpha} \propto -q^{3}\ln q$ is observed near $q_{\alpha_\text{BP}}$; however, this logarithmic scaling terminates at $q_{\alpha_\text{0}}$.
}
\end{figure}

\subsection{Two-dimensional system}
We also study a 2D amorphous system model (see Section~\ref{system}).
The results are presented in Figs.~\ref{fig1}b,~\ref{fig2}b,~\ref{fig3}b,~\ref{fig6}, and~\ref{figs10}.
The phonon transport modes share similar properties between the 2D and 3D cases.
In the BP regime, $\Omega \sim \omega_\text{BP}$, we see a dip in the sound speed $c_\alpha$ and the $\Omega^2$ law, $\Gamma_\alpha \propto \Omega^{2}$~(see Fig.~\ref{fig1}b).
In the continuum limit, $\Omega \lesssim \omega_0$, characteristics of Rayleigh scattering behavior, $c_\alpha = c_{\alpha 0}$ and $\Gamma_\alpha \propto \Omega^3$, are observed.
We also obtain similar laws for the scaling with the pressure $p$ as in the 3D case~(see Fig.~\ref{fig2}b):
\begin{equation} \label{cscaling2d}
\begin{aligned}
c_\alpha(\Omega) = 
\left\{ \begin{aligned}
& c_{\alpha 0} f_{\alpha \text{min}} & (\Omega \sim \omega_\text{BP}), \\
& c_{\alpha 0} & (\Omega \lesssim \omega_\text{0}),
\end{aligned} \right.
\end{aligned} 
\end{equation}
\begin{equation} \label{gammascaling2d}
\begin{aligned}
\Gamma_\alpha(\Omega) = 
\left\{ \begin{aligned}
& B_{\alpha \text{BP}} \omega_\ast^{n_\alpha} \left( \frac{\Omega}{\omega_\ast} \right)^2 & (\Omega \sim \omega_\text{BP}), \\
& B_{\alpha 0} \omega_\ast^{n_\alpha} \left( \frac{\Omega}{\omega_\ast} \right)^3 & (\Omega \lesssim \omega_\text{0}),
\end{aligned} \right. 
\end{aligned}
\end{equation}
with the exponents of $n_T = 1.0$ (transverse waves) and $n_L = 1.6$ (longitudinal waves).
Here $f_{T \text{min}} = 0.97$, $f_{L \text{min}} = 0.99$, $B_{T \text{BP}} = 1.3$, $B_{L \text{BP}}=0.80$, $B_{T 0} = 17$, and $B_{L 0}=7.5$.

A notable difference from the 3D case is that the vDOS of the generalized Debye model converges to the Debye level at $\omega \lesssim \omega_0$~(see Fig.~\ref{fig3}b), as do the numerical values found through vibrational mode analysis~\cite{Mizuno_2017}.
This difference is understood by the fact that $g_1(\omega)$ is much larger than $g_2(\omega)$ at $\omega \lesssim \omega_0$ in the 2D case.
This is then rationalized by considering that $g_1(\omega)$ is dominated by the Debye vDOS that is a linear function of $\omega/\omega_\ast$, i.e., the phonon modes are much more abundant in 2D than in 3D.
This also means that the total vDOS is expressed as a function of $\omega/\omega_\ast$, as indeed demonstrated in Fig.~\ref{fig6}.
Interestingly, although the direct analysis of vibrational eigenmodes cannot separate the non-phonon modes from the phonon modes in 2D~\cite{Mizuno_2017}, the generalized Debye model enables us to determine the nonphonon vDOS $g_\text{nonphonon}(\omega)$ as $g_2(\omega)$, whose $\omega$ and $p$ dependences are controlled by the term $\Gamma_T/c_{T0}^2$ (see Eq.~(\ref{2dpvdos})):
\begin{equation}
\begin{aligned} \label{vdosgamma2d}
& g_\text{nonphonon}(\omega) \approx \frac{1}{\pi q_D^2} \frac{\Gamma_T (\Omega = \omega)}{c_{T0}^2} \log\left( \frac{c_{T0}^2 q_D^2}{\omega^2} \right), \\
& \quad \propto
\left\{ \begin{aligned}
& \left( \frac{\omega}{\omega_\ast} \right)^2 \log \left( \frac{\omega_\ast}{\omega^2} \right) \propto \left( \frac{\omega}{\omega_\ast} \right)^2 & (\omega \sim \omega_\text{BP}), \\
& \left( \frac{\omega}{\omega_\ast} \right)^3 \log \left( \frac{\omega_\ast}{\omega^2} \right) \propto \left( \frac{\omega}{\omega_\ast} \right)^3 & (\omega \lesssim \omega_\text{0}).
\end{aligned} \right.
\end{aligned}
\end{equation}
Figure~\ref{figs10} plots $g_2(\omega) \approx g_\text{nonphonon}(\omega)$ and demonstrates Eq.~(\ref{vdosgamma2d}).

\subsection{Logarithmic scaling of sound attenuation}
The recent work~\cite{Gelin_2016} suggested a logarithmic scaling of $\Gamma_\alpha \propto -q^{d+1} \ln q$ with the wavenumber $q$ and questioned the validity of the Rayleigh scattering law $\Gamma_\alpha \propto q^{d+1}$.
They also argued that the logarithmic correction to the Rayleigh law might originate from the long-range correlation of elastic modulus~\cite{DeGiuli_2018}.
To elucidate this point, we perform additional simulations in 2D at the pressure $p=5 \times 10^{-2}$, by using larger systems of up to $N=4096000$.
Figure~\ref{figs23} shows $\Gamma_\alpha/q^3$ against $q$ in a semilog plot.
We can find that the suggested logarithmic scaling works in the BP regime, $q \sim q_{\alpha \text{BP}} \equiv \omega_\text{BP}/c_{\alpha 0}$; however, this scaling becomes invalid at the wavenumber corresponding to the continuum limit, $q \lesssim q_{\alpha \text{0}} \equiv \omega_\text{0}/c_{\alpha 0}$, where $\Gamma_\alpha \propto q^{3}$ clearly appears.
We therefore demonstrate that with decreasing $q$ (or $\Omega$), the Rayleigh scattering law emerges in the continuum limit.
It is important to emphasize that these behaviors are not affected by the system size (data from different system sizes of $N=1024000$ to $4096000$ coincide within error bars).
We also analyze the unstressed systems (in both 2D and 3D) and plot $\Gamma_\alpha / q^{d+1}$ against $q$ in Fig.~S10 of the Supplemental Material~\cite{supplement}.
We can cleanly observe the Rayleigh scattering law without logarithmic correction in the low wavenumber regime, $q < q_{\alpha 0}$.
We therefore conclude that the Rayleigh law is valid in the 2D system and the 3D and 2D unstressed systems, and expect that it can be also valid in the 3D system.

In addition, we study the spatial correlations in stress field and (affine) elastic modulus field (see Appendix~\ref{moduluscorrelation}) and plot the results in Fig.~\ref{figs92}.
The present amorphous solid shows the long-range correlation in stress field, but not any long-range correlation in elastic modulus field, which is in contrast to the system studied in Ref.~\cite{Gelin_2016}.
We also confirm no long-range correlation in elastic modulus field in the unstressed systems (see Fig.~S13 of the Supplemental Material~\cite{supplement}).
Our results therefore do not exclude the possible relation between the logarithmic correction to the Rayleigh law and the long-range nature of elastic modulus, as proposed by Ref.~\cite{Gelin_2016}.

\section{Conclusion}
By means of large-scale numerical simulations, we achieved a consistent understanding of the phonon transport and vibrational eigenmodes in amorphous solids.
Near the BP, at $\Omega \sim \omega_\text{BP}$, $\Omega^2$ law (with $\Gamma_\alpha \propto \Omega^2$) is observed, which is linked to the disordered and extended nature of the eigenmodes.
In the continuum limit, at $\Omega < \omega_\text{ex0}$, Rayleigh scattering (with $c_\alpha = c_{\alpha 0}$ and $\Gamma_\alpha \propto \Omega^{4}$) is observed, which is linked to a mixture of phonon modes and soft localized modes.
In this regime, the soft localized modes play the role of defects to enhance the phonon scattering.
The crossover frequency is therefore identified as $\omega_\text{ex0}$.
Our results also unambiguously demonstrate the occurrence of Rayleigh scattering without logarithmic correction and thus shed new light regarding the argument for a logarithmic scaling law presented in the recent work~\cite{Gelin_2016}.
We would argue that the Rayleigh law is valid in the present amorphous solid; however, it might be altered by the long-range nature of (affine) elastic modulus which is absent in the present system.

We also established the jamming scaling laws for the phonon transport properties.
We find that the $c_\alpha$ and $\Gamma_\alpha$ values measured over a wide range of pressures can be seen to collapse when they are scaled properly by $\omega_\ast \propto p^{1/2}$.
Based on these results, we reveal the length scale $D_\alpha$ of the sources responsible for the Rayleigh scattering.
The scaling behaviors of $D_\alpha$ coincide with those of the length scales characterizing the continuum limits of elastic response and phonon transport.
This result suggests that all of these length scales are controlled by the correlation length scale $\ell_c$ of the vibrational eigenmodes at $\omega_\ast$~\cite{Lerner_2014,Yan_2016}.
In addition, by applying the generalized Debye model, we find that the non-Debye laws for nonphonon modes are intimately related to the phonon scattering laws in the form, $g_\text{nonphonon}(\omega) \propto \Gamma_T/c_{T0}^2$: $g_\text{nonphonon}(\omega) \propto (\omega/\omega_\ast)^4$ at $\omega < \omega_\text{ex0}$ and $\propto (\omega/\omega_\ast)^2$ at $\omega \sim \omega_\text{BP}$.

The results of the jamming scaling laws enabled us to directly test the mean-field theory of jamming.
We find that in the BP regime, the jamming scaling laws are consistent between our simulation results and the predictions of effective medium theory~\cite{Wyart_2010,DeGiuli_2014}.
However, we find inconsistencies in the continuum limit regime, with the theory significantly underestimating the strength of the Rayleigh scattering.
One important issue is that the effect of the soft localized modes is neglected in the theory.
Nevertheless, we also notably show that the theory does not correctly describe the phonon scattering even in the unstressed system, where the soft localized modes are strongly suppressed~\cite{Mizuno_2017,Lerner2_2017}.
Our results therefore reveal crucial issues that must be solved with regard to the current version of the theory~\cite{Wyart_2010,DeGiuli_2014}.

Here we make some remarks on the other theoretical approaches; the elastic heterogeneities, and the two-level system and the soft potential model.
The mean field theory of elastic heterogeneities~\cite{schirmacher_2006,schirmacher_2007,Marruzzo_2013,Schirmacher_2015} uses the effective medium technique to predict the Rayleigh scattering and $\Omega^2$ law, as does the mean-field theory of jamming~\cite{Wyart_2010,DeGiuli_2014}.
Since effect of the soft localized modes is also missing in the theory of elastic heterogeneities, this theory has the same issue as the mean-field theory of jamming; the phonon scattering is significantly underestimated.
For the two-level system and the soft potential model~\cite{Anderson_1972,Karpov_1983,Buchenau_1991,Buchenau_1992,Gurevich_2003,Gurevich_2005,Gurevich_2007}, it could be an important subject to clarify anharmonic nature of the vibrational properties that the mean-field theories of jamming and elastic heterogeneities do not take care of.
Anharmonicities are expected to play crucial roles in the low-temperature thermal properties below $1$~[K]~\cite{lowtem,Zeller_1971,Graebner_1986}.

Finally, we mention the hybridization effect due to phonon broadening.
In a recent work~\cite{Bouchbinder_2018}, it was reported that the soft localized modes hybridize with the phonon modes as the system size increases. 
The size of our investigated system in 3D is still too small for us to observe this effect directly.
Yet, we speculate that this hybridization does not exert a qualitative influence on the phonon transport, and the crossover frequency $\omega_\text{ex0}$ can be determined through the phonon transport.
Indeed, the Rayleigh scattering has been observed in many experimental works and should be free of system size effects~\cite{ruffle_2006,Masciovecchio_2006,Monaco_2009,Baldi_2010,Baldi_2011,Baldi_2016}.
Also, although the soft localized modes strongly hybridize with the phonon modes in our 2D case, we still observe the Rayleigh scattering and can determine the crossover frequency $\omega_0$ (Figs.~\ref{fig1}b and \ref{fig2}b).
These observations imply that the phonon transport does not qualitatively change due to the hybridization.
We therefore speculate that even if the hybridization takes place, we can determine the frequency $\omega_\text{ex0}$ in 3D at which vibrational properties (vibrational eigenmodes as well as phonon transport) show the crossover.
However, it will certainly be important to investigate the effects of hybridization in future work.

\section*{Acknowledgments}
We thank M.~Wyart, E.~DeGiuli, B.~P.~Tighe, A.~Zaccone, C.~Caroli, and A.~Lemaitre for useful discussions and suggestions.
This work was supported by a Grant-in-Aid for Young Scientists B (No.~17K14369), a Grant-in-Aid for Young Scientists A (No.~17H04853), and a Grant-in-Aid for Scientific Research B (No.~16H04034) from the Japan Society for the Promotion of Science (JSPS).
The theoretical calculations were partially performed using the Research Center for Computational Science, Okazaki, Japan.

\begin{figure*}[t]
\centering
\includegraphics[width=0.95\textwidth]{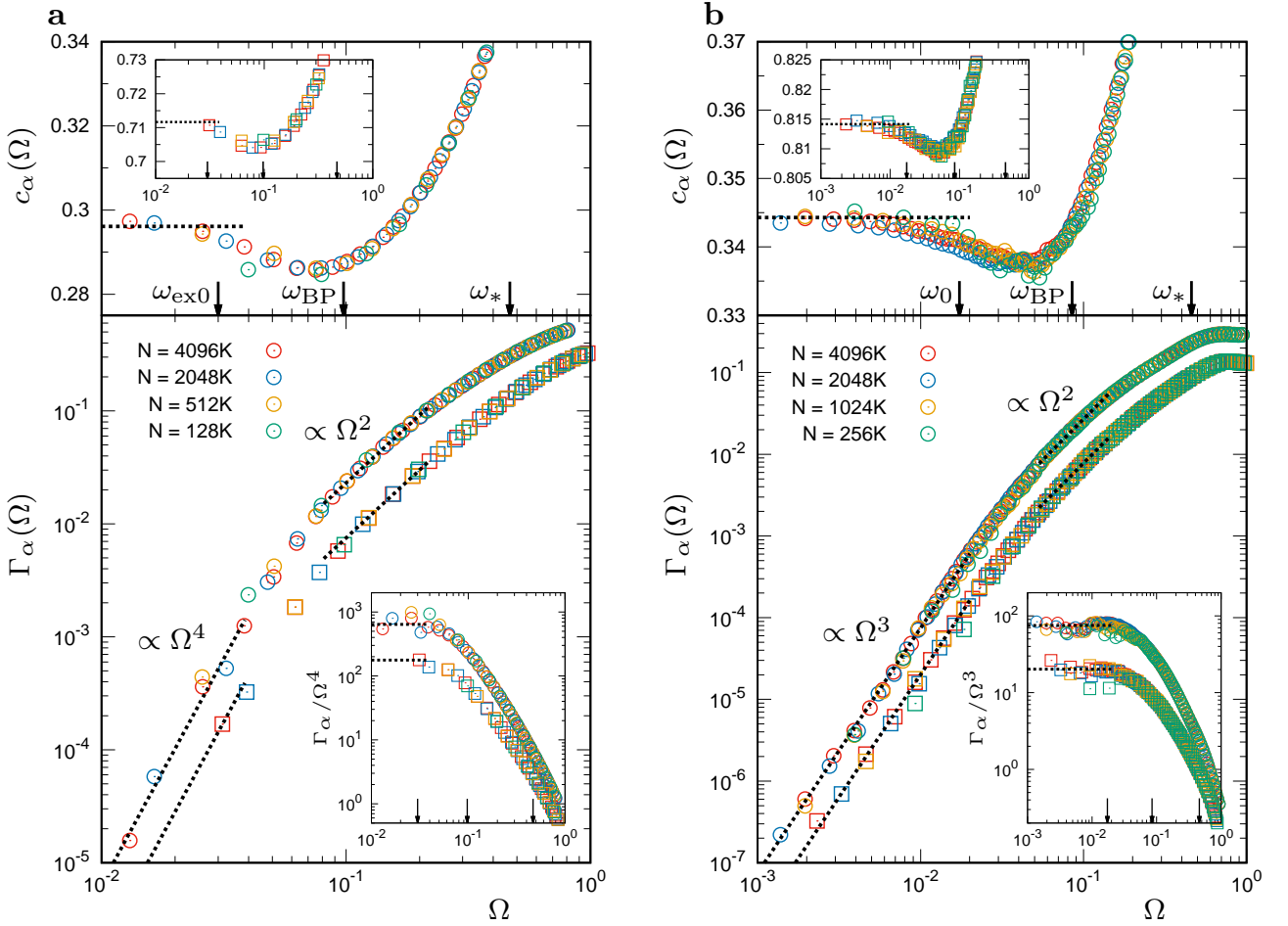}
\vspace*{0mm}
\caption{\label{figs7}
{Effects of system size on phonon transport.}
Plots of the sound speed $c_\alpha (\Omega)$ and the attenuation rate $\Gamma_\alpha (\Omega)$ as functions of $\Omega$, for transverse ($\alpha =T$, circles) and longitudinal ($\alpha = L$, squares) waves.
{\bf a} (left panels), The 3D model system ($d=3$).
{\bf b} (right panels), The 2D model system ($d=2$).
The packing pressure is $p=5 \times 10^{-2}$.
Data are plotted for different system sizes of $N=128000$, $512000$, $2048000$, and $4096000$ for the 3D case and $N=256000$, $1024000$, $2048000$, and $4096000$ for the 2D case.
The data for $N=2048000$ and $4096000$ (3D) and for $N=1024000$ (2D) are the same as those presented in Fig.~\ref{fig1}.
We confirm no apparent differences among systems of different sizes.
}
\end{figure*}

\begin{figure*}[t]
\centering
\includegraphics[width=0.95\textwidth]{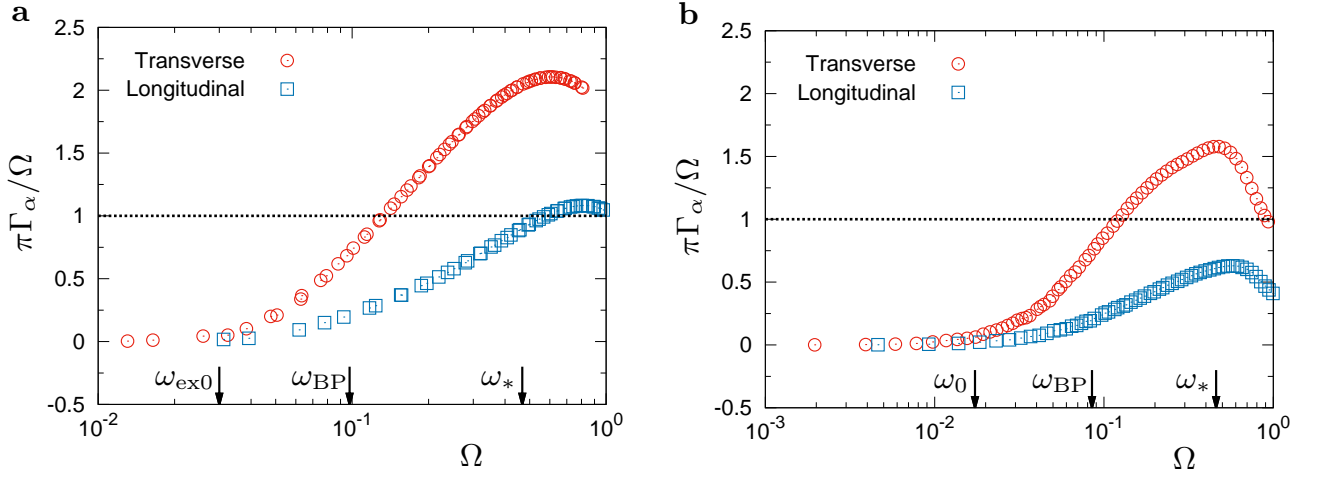}
\vspace*{0mm}
\caption{\label{figs5}
{The Ioffe-Regel (IR) frequency of phonon transport.}
The ratio $\pi \Gamma_\alpha/\Omega$ is plotted as a function of $\Omega$, for transverse ($\alpha =T$, red circles) and longitudinal ($\alpha = L$, blue squares) waves.
{\bf a}, The 3D model system ($d=3$).
{\bf b}, The 2D model system ($d=2$).
The packing pressure is $p=5 \times 10^{-2}$.
The frequency at which this ratio is equal to one ($\pi \Gamma_\alpha/\Omega = 1$) is defined as the IR frequency, $\Omega_{\alpha \text{IR}}$.
Figure~\ref{figs6a} plots the dependence of $\Omega_{\alpha \text{IR}}$ on the pressure $p$.
}
\end{figure*}

\begin{figure*}[t]
\centering
\includegraphics[width=0.95\textwidth]{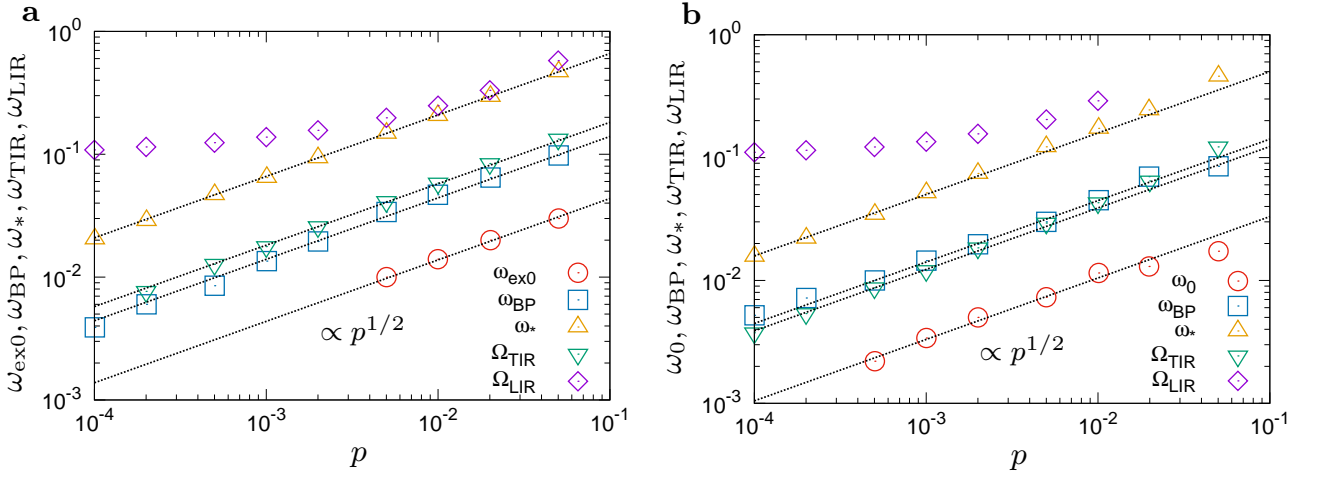}
\vspace*{0mm}
\caption{\label{figs6a}
{Packing pressure dependences of the characteristic frequencies.}
Plots of $\omega_{\text{ex}0}$ ($\omega_0$), $\omega_\text{BP}$, $\omega_\ast$, $\Omega_\text{TIR}$, and $\Omega_\text{LIR}$ as functions of $p$.
{\bf a}, The 3D model system ($d=3$).
{\bf b}, The 2D model system ($d=2$).
The data for $\omega_{\text{ex}0}$ ($\omega_0$), $\omega_\text{BP}$, and $\omega_\ast$ are the same as those presented in Ref.~\cite{Mizuno_2017}.
The lines represent the power-law scaling $\propto p^{1/2}$.
For all pressures $p$, the IR frequency $\Omega_\text{TIR}$ for transverse waves almost coincides with the BP frequency, $\Omega_\text{TIR} \approx \omega_\text{BP}$.
On the other hand, the IR frequency $\Omega_\text{LIR}$ for longitudinal waves is much larger than $\Omega_\text{TIR} \approx \omega_\text{BP}$.
}
\end{figure*}

\begin{figure}[t]
\centering
\includegraphics[width=0.475\textwidth]{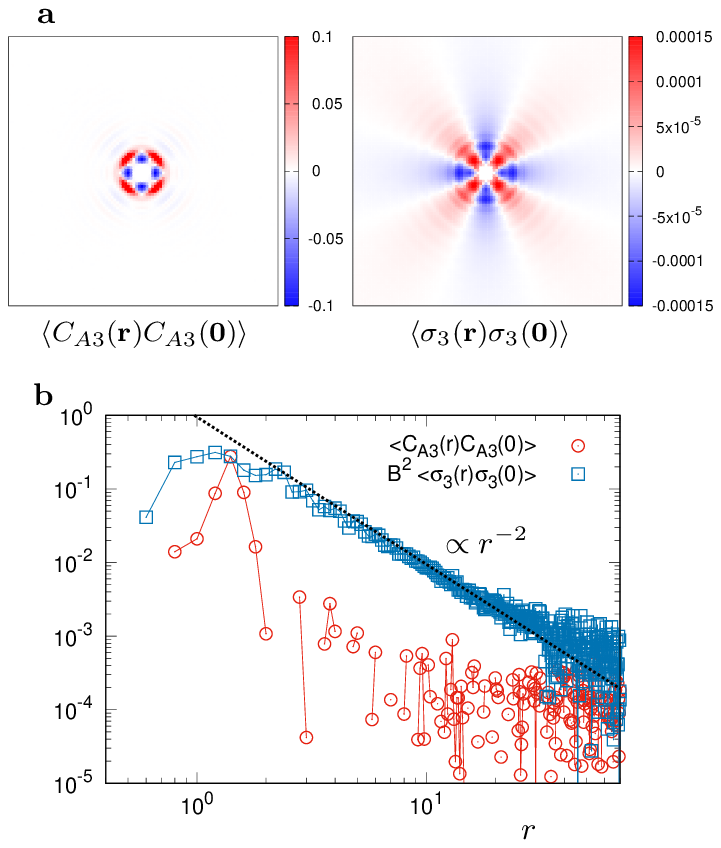}
\vspace*{0mm}
\caption{\label{figs92}
{Spatial correlations of elastic modulus and stress in the 2D amorphous solid.}
{\bf a}, Spatial autocorrelation functions of the elastic modulus field $C_{A3}(\mathbf{r})$ and the stress field $\sigma_3(\mathbf{r})$.
{\bf b}, Cut of the correlation functions along the $\pi/4$ axis.
The packing pressure is $p=5 \times 10^{-2}$.
In {\bf b}, the spatial correlation of the stress shows a clear $r^{-2}$ scaling, while the elastic modulus does not.
}
\end{figure}

\appendix

\section{System size effects}~\label{sizesection}
To study the system size effects, we analyzed additional system sizes of $N=128000$ and $512000$ for the 3D case and $N=256000$, $2048000$, and $4096000$ for the 2D case, at the packing pressure $p=5 \times 10^{-2}$.
Figure~\ref{figs7} compares the results for different system sizes and confirms no apparent differences among them (see also Fig.~\ref{figs23} for the 2D case).
We therefore conclude that no apparent system size effects appear in our results.

\section{The Ioffe-Regel (IR) frequency}~\label{IRsection}
We define the IR frequency $\Omega_{\alpha \text{IR}}$ ($\alpha = T$ or $L$) as $\pi \Gamma_\alpha(\Omega_{\alpha \text{IR}}) / \Omega_{\alpha \text{IR}} = 1$, where $\Gamma_\alpha (\Omega)$ is considered a function of $\Omega$~(see Fig.~\ref{figs5}).
Above $\Omega = \Omega_{\alpha \text{IR}}$, the phonon decay time ($= \Gamma_\alpha^{-1}$) becomes shorter than half of the vibrational period ($= \pi/\Omega$); i.e., the phonon decays within half of the duration of one period.
The IR frequency $\Omega_{\alpha \text{IR}}$ therefore corresponds to an upper bound on the propagation frequency of a phonon as a plane wave.

Figure~\ref{figs6a} plots $\Omega_{\alpha \text{IR}}$ as a function of the pressure $p$.
At any $p$, $\Omega_{\text{TIR}}$ for transverse waves coincides with the BP frequency, $\Omega_\text{TIR} \approx \omega_\text{BP} \propto p^{1/2}$.
This coincidence has been also observed in other glassy systems~\cite{ruffle_2006,shintani_2008,Monaco2_2009,Mizuno_2014}.
Note that $\Omega_{\text{LIR}}$ of the longitudinal waves is located at the higher frequency, $\Omega_\text{LIR} \gg \Omega_\text{TIR} \approx \omega_\text{BP}$, that has been reported for a similar system as the present one~\cite{Wang_2015}.
We therefore conclude that the phonon persists as a plane wave below the BP and that the phonon approximation, $\Omega \gg \Gamma_T$, which is assumed in the generalized Debye model in Section~\ref{sectionGDM}, is valid for $\Omega \lesssim \omega_\text{BP}$.

It is worth to note that the fitting function in Eq.~(\ref{ctfunction}) (i.e., the damped harmonic oscillator model) may not be appropriate to measure the phonon transport properties in the frequency regime above the IR frequency~\cite{Beltukov_2018}.
A recent work~\cite{Damart_2017} proposed a theoretical expression to measure the attenuation rate by using the mechanical spectroscopy.
Here we do not go into this direction more, because the present work mainly focuses on the frequency regime $\Omega \lesssim \omega_\text{BP}$, i.e., below the IR frequency, where we can safely use the fitting function in Eq.~(\ref{ctfunction}).

\section{Spatial correlations of elastic modulus and stress fields}~\label{moduluscorrelation}
Following Ref.~\cite{Gelin_2016}, we study the spatial correlations of the (affine) elastic modulus and the stress for the 2D case.
We define the particle-based, affine elastic modulus $C_{A\alpha}$ ($\alpha = 1,2,3,4,5$) for the particle $i$ as
\begin{equation}
\begin{aligned}
C_{A1}(\mathbf{r}_{i}) &= \sum_{j \in \partial_i} h_{ij}, \\
C_{A2}(\mathbf{r}_{i}) &= \sum_{j \in \partial_i} h_{ij} \cos(2 \theta_{ij}), \\
C_{A3}(\mathbf{r}_{i}) &= \sum_{j \in \partial_i} h_{ij} \sin(2 \theta_{ij}), \\
C_{A4}(\mathbf{r}_{i}) &= \sum_{j \in \partial_i} h_{ij} \cos(4 \theta_{ij}), \\
C_{A5}(\mathbf{r}_{i}) &= \sum_{j \in \partial_i} h_{ij} \sin(4 \theta_{ij}),
\end{aligned}
\end{equation}
where $h_{ij} = \phi''(r_{ij}) {r_{ij}}^2 - \phi'(r_{ij})r_{ij}$ with $r_{ij}$ the distance between particles $i$ and $j$, $\mathbf{r}_{ij}/r_{ij} = (\cos \theta_{ij},\sin \theta_{ij})$, $\mathbf{r}_{ij} = \mathbf{r}_{j} - \mathbf{r}_{i}$, and $\sum_{j \in \partial_i}$ denotes the summation of particles $j$ that interact with the particle $i$.
We note that $C_{A \alpha}$ is the affine modulus, not including the non-affine component~\cite{Mizuno3_2016}.
$C_{A1}/4$ corresponds to the bulk modulus, and $(C_{A1}+C_{A4})/8$ and $(C_{A1}-C_{A4})/16$ are two shear moduli.

We also define the particle-based stress $\sigma_{\alpha}$ ($\alpha = 1,2,3$) for the particle $i$ as
\begin{equation}
\begin{aligned}
\sigma_1(\mathbf{r}_{i}) &= \sum_{j \in \partial_i} p_{ij}, \\
\sigma_2(\mathbf{r}_{i}) &= \sum_{j \in \partial_i} -p_{ij} \cos(2 \theta_{ij}), \\
\sigma_3(\mathbf{r}_{i}) &= \sum_{j \in \partial_i} -p_{ij} \sin(2 \theta_{ij}),
\end{aligned}
\end{equation}
where $p_{ij} = - (1/2) \phi'(r_{ij})r_{ij}$.
$\sigma_1$ corresponds to the pressure, and $\sigma_2$ and $\sigma_3$ are two shear stresses.
We then calculate the spatial autocorrelation functions of the elastic modulus field $C_{A\alpha}(\mathbf{r})$ and the stress field $\sigma_\alpha(\mathbf{r})$.

Figure~\ref{figs92} plots the autocorrelation functions, $\left<C_{A3}(\mathbf{r}) C_{A3}(\mathbf{0}) \right>$ and $B^2 \left<\sigma_3(\mathbf{r}) \sigma_3(\mathbf{0}) \right>$, at the packing pressure $p=5 \times 10^{-2}$.
Here $B=4K_A/P$ is the ratio of the global affine bulk modulus $K_A = (1/2L^{2})\sum_i C_{A1}(\mathbf{r}_{i})/4$ and the pressure $P = (1/2L^{2})\sum_i \sigma_1(\mathbf{r}_{i})$~\cite{Gelin_2016}.
The stress field shows the power-law correlation $\propto r^{-2}$, while the elastic modulus field does not show such the long-range correlation.
This situation is different from that in Ref.~\cite{Gelin_2016}.
Because the interparticle potential studied in Ref.~\cite{Gelin_2016} is the inverse-power-law and $h_{ij} \propto p_{ij}$, the power-law correlation in stress causes the analogous correlation in elastic modulus as $\left<C_{A3}(\mathbf{r}) C_{A3}(\mathbf{0}) \right> \propto \left<\sigma_3(\mathbf{r}) \sigma_3(\mathbf{0}) \right> \propto r^{-2}$.
On the other hand, the interparticle potential of the present system does not show such the property: the stress shows the long-range correlation, but it does not induce the analogous correlation in the elastic modulus, as is demonstrated in Fig.~\ref{figs92}.
We also confirm that the unstressed system shows almost the same spatial correlation in elastic modulus, as the original stressed system (see Fig.~S13).

\section{Scattering theory of an elastic wave}~\label{scattering}
Here, we review the scattering theory of an elastic wave in a three-dimensional space~\cite{Bhatia_1967}.
Our goal here is to evaluate the rate of attenuation due to the presence of scattering sources, i.e., elastic inhomogeneities.
Let us consider an elastic wave propagating in an isotropic elastic medium embedded with a single scattering source.
The displacement field $\mathbf{s}(\mathbf{r},t)$ (where $\mathbf{r}$ is a spatial vector and $t$ is time) is written as $\mathbf{s}(\mathbf{r},t) = \mathbf{u}(\mathbf{r}) e^{\mathrm{i} \Omega t}$, where $\Omega$ is the propagation frequency.
We consider the inhomogeneity in the elastic modulus as the scattering source. The modulus tensor is described as a sum of two terms, $C_{ijkl} + \delta C_{ijkl}(\mathbf{r})$.
$C_{ijkl} = ( K -2G/3 ) \delta_{ij} \delta_{kl} + G ( \delta_{ik}\delta_{jl} +\delta_{il}\delta_{jk})$ is the isotropic modulus tensor with the bulk modulus $K$ and the shear modulus $G$.
$\delta C_{ijkl}(\mathbf{r})$ is the elastic inhomogeneity; it takes a constant, non-zero value at $\mathbf{r} \in V$, and otherwise, $\delta C_{ijkl}(\mathbf{r}) =0$, where $V$ denotes the volume of the scattering source.
We start with the equation of motion:
\begin{equation}
\begin{aligned} \label{waveeq1}
& (C_{ijkl} + \delta C_{ijkl}) \frac{\partial^2 u_k}{\partial x_j \partial x_l} + \rho \Omega^2 u_i = 0,
\end{aligned}
\end{equation}
where $\rho$ is the mass density of the elastic medium, which is assumed to be spatially uniform, $\mathbf{r}=(x,y,z)=(x_1,x_2,x_3)$, and the Einstein summation convention is employed.

We suppose that the displacement $\mathbf{u}(\mathbf{r})$ can be decomposed into components corresponding to the incident wave and the scattered wave: $\mathbf{u}(\mathbf{r}) = \mathbf{u}^\text{in}(\mathbf{r}) + \mathbf{u}^\text{sc}(\mathbf{r})$.
In addition, we consider that the incident wave propagates in the $z$ direction and that its wave vector is $\mathbf{q}^\text{in}=q \hat{\mathbf{q}}^\text{in} = q(0,0,1)$: $\mathbf{u}^\text{in} (\mathbf{r}) = {\mathbf{a}}^\text{in} e^{-\mathrm{i} \mathbf{q}^\text{in}\cdot \mathbf{r} } = {\mathbf{a}^\text{in}} e^{-\mathrm{i} qz}$.
${\mathbf{a}}^\text{in}$ is the polarization vector, where $\mathbf{a}^\text{in}={\mathbf{a}}^\text{in}_T = (a_x,a_y,0)$ for a transverse wave and $\mathbf{a}^\text{in}={\mathbf{a}}^\text{in}_L = (0,0,a_z)$ for a longitudinal wave.
The equation of motion for the incident wave is
\begin{equation}
\begin{aligned} \label{waveeq2}
& C_{ijkl} \frac{\partial^2 u^\text{in}_k}{\partial x_j \partial x_l} + \rho \Omega^2 u^\text{in}_i = 0,
\end{aligned}
\end{equation}
which yields a linear dispersion relation, $\Omega/q = c_{T0} = \sqrt{G/\rho}$ for a transverse wave and $\Omega/q = c_{L0} = \sqrt{(K+4G/3)/\rho}$ for a longitudinal wave.
We can therefore calculate the incident energy per unit time and area for a transverse wave ($\alpha=T$) or a longitudinal wave ($\alpha=L$) as follows:
\begin{equation} \label{waveeq3}
E^\text{in}_\alpha = c_{\alpha 0} \rho \Omega^2 \left| \mathbf{s}^\text{in}(\mathbf{r},t) \right|^2 = \rho c_{\alpha 0}^3 q^2 \left| {\mathbf{a}}^\text{in}_\alpha \right|^2.
\end{equation}

We next assume that $\delta C_{ijkl} / C_{ijkl} \ll 1$ and $|\mathbf{u}^\text{sc}|/|\mathbf{u}^\text{in}| \ll 1$ in Eq.~(\ref{waveeq1}) and obtain the equation of motion for the scattered wave $\mathbf{u}^\text{sc}(\mathbf{r})$~\cite{Bhatia_1967}:
\begin{equation}
\begin{aligned} \label{waveeq4}
& C_{ijkl} \frac{\partial^2 u^\text{sc}_k}{\partial x_j \partial x_l} + \rho \Omega^2 u^\text{sc}_i = - \delta C_{ijkl} \frac{\partial^2 u^\text{in}_k}{\partial x_j \partial x_l}.
\end{aligned}
\end{equation}
We denote the wave vector of the scattered wave by $\mathbf{q}^\text{sc} = q \hat{\mathbf{q}}^\text{sc}$ ($\hat{\mathbf{q}}^\text{sc}$ is a unit vector).
Then, the scattered wave can be described as $\mathbf{u}^\text{sc}(\mathbf{r}) \propto e^{- \mathrm{i} q \hat{\mathbf{q}}^\text{sc} \cdot \mathbf{r}}$ and can be decomposed into a transverse component $\mathbf{u}^\text{sc}_T$ and a longitudinal component $\mathbf{u}^\text{sc}_L$, as follows: $\mathbf{u}^\text{sc}_T = \mathbf{u}^\text{sc} - \mathbf{u}^\text{sc}_L$ and $\mathbf{u}^\text{sc}_L = (\hat{\mathbf{q}}^\text{sc} \cdot \mathbf{u}^\text{sc})\hat{\mathbf{q}}^\text{sc}$.
Here, we introduce the quantities $\mathbf{\Psi}^\text{sc} = \nabla \times \mathbf{u}^\text{sc}$ and $\Delta^\text{sc} = \nabla \cdot \mathbf{u}^\text{sc}$, which are related to $\mathbf{u}^\text{sc}_T$ and $\mathbf{u}^\text{sc}_L$ as follows: $\left| \mathbf{u}^\text{sc}_T \right| = \left| \mathbf{\Psi}^\text{sc} \right| /q$ and $\left| \mathbf{u}^\text{sc}_L \right| = \left| {\Delta^\text{sc}} \right|/q$.
From Eq.~(\ref{waveeq4}), we obtain the equations for $\mathbf{\Psi}^\text{sc}$ and $\Delta^\text{sc}$:
\begin{equation}
\begin{aligned} \label{waveeq5}
&  \nabla^2 \mathbf{\Psi}^\text{sc} + \left( \frac{\Omega}{c_{T0}} \right)^2 \mathbf{\Psi}^\text{sc} = \nabla \times \left( {\mathbf{a}^\text{in}} q^2 \delta \gamma_T e^{-\mathrm{i} qz}\right), \\
&  \nabla^2 {\Delta}^\text{sc} + \left( \frac{\Omega}{c_{L0}} \right)^2 {\Delta}^\text{sc} = \nabla \cdot \left( {\mathbf{a}^\text{in}} q^2 \delta \gamma_L e^{-\mathrm{i} qz}\right),
\end{aligned}
\end{equation}
where $\delta \gamma_T = \delta G / G$ and $\delta \gamma_L = \delta (K + 4G/3) / (K + 4G/3)$ represent the strengths of the scattering source (inhomogeneity) for transverse and longitudinal waves, respectively.
To derive Eq.~(\ref{waveeq5}), we assume that $\delta C_{xzzz}=\delta C_{yzzz}=0$ and $\delta C_{yzxz}=\delta C_{zzxz}=\delta C_{xzyz}=\delta C_{zzyz}=0$.
We also remark that transverse and longitudinal waves are scattered by different elastic inhomogeneities, $\delta G(\mathbf{r})$ and $\delta (K + 4G/3)(\mathbf{r})$, respectively.
The volumes of these scattering sources are generally different, and we use $V_T$ to denote the volume of $\delta G(\mathbf{r})$ and $V_L$ to denote the volume of $\delta (K + 4G/3)(\mathbf{r})$.

The particular solution to Eq.~(\ref{waveeq5}) is an outward-going spherical wave~\cite{Bhatia_1967}.
For a wave with a wavelength $\lambda = 2\pi / q$ that is much larger than the length scale of the scattering source, $D = V^{1/3}$, we obtain the following solution:
\begin{equation}
\begin{aligned} \label{waveeq6}
\mathbf{\Psi}^\text{sc} (\mathbf{r}) & \approx  \left( \mathrm{i} q^3 \hat{\mathbf{q}}^\text{in} \times {\mathbf{a}}^\text{in} \right) \delta \gamma_T V_T \frac{e^{-\mathrm{i}q r}}{4 \pi r},\\
{\Delta}^\text{sc} (\mathbf{r}) & \approx  \left( \mathrm{i} q^3 \hat{\mathbf{q}}^\text{in} \cdot {\mathbf{a}}^\text{in} \right) \delta \gamma_L V_L  \frac{e^{-\mathrm{i}q r}}{4 \pi r}.
\end{aligned}
\end{equation}
We finally obtain the energy scattered in all directions per unit time as follows:
\begin{equation} \label{waveeq7}
\begin{aligned}
E^\text{sc}_\alpha &= c_{\alpha 0} \rho \Omega^2 \int \left| \mathbf{s}^\text{sc}(\mathbf{r},t) \right|^2 r^2 d\Omega, \\
 &= \rho c_{\alpha 0}^3 q^2 \left| {\mathbf{a}}^\text{in}_\alpha \right|^2 \frac{ \delta \gamma_\alpha^2}{4 \pi} V_\alpha^2 q^4 = E^\text{in}_\alpha \frac{ \delta \gamma_\alpha^2}{4 \pi} V_\alpha^2 q^4,
\end{aligned}
\end{equation}
where $d\Omega$ is the solid angle.
The scattering cross section $\sigma^\text{tot}_\alpha$ is therefore calculated as the ratio between the scattered energy and the incident energy:
\begin{equation} \label{waveeq8}
\sigma^\text{tot}_\alpha \equiv \frac{E^\text{sc}_\alpha}{E^\text{in}_\alpha} = \frac{\delta \gamma_\alpha^2}{4 \pi} V_\alpha^2 q^4.
\end{equation}
This demonstrates the occurrence of Rayleigh scattering behavior, $\sigma^\text{tot}_\alpha \propto q^4$, in the long-wavelength and low-frequency regime.

Up to now, we have studied the scattering of an elastic wave by a single scattering source.
In an amorphous solid, however, many scattering sources should be considered to scatter an elastic wave.
Here, we assume that the number density of these scattering sources is $n_\alpha = V_\alpha^{-1}$, as in the case of polycrystalline solids~\cite{Bhatia_1967,West_1984}, and that these scattering sources independently scatter the elastic wave.
We can then formulate the fraction of the scattered energy per unit time, i.e., the attenuation rate, as follows:
\begin{equation} \label{waveeq9}
\begin{aligned}
\Gamma_\alpha &= \sigma^\text{tot}_\alpha n_\alpha c_{\alpha 0} = \frac{\delta \gamma_\alpha^2}{4 \pi} V_\alpha q^4 c_{\alpha 0}, \\
 &= \frac{\delta \gamma_\alpha^2}{4 \pi} \left( \frac{D_\alpha}{c_{\alpha 0}} \right)^3 \Omega^4 \propto \left( \frac{D_\alpha}{c_{\alpha 0}} \right)^3 \Omega^4.
\end{aligned}
\end{equation}
%

\bibliographystyle{apsrev4-1}
\bibliography{reference}

\end{document}